\documentclass[11pt]{article}
\usepackage{jheppub}
\usepackage[utf8]{inputenc}
\usepackage{amsmath,amssymb,amsthm,mathtools,bm}
\usepackage{xcolor}
\usepackage{hyperref}
\hypersetup{colorlinks=true,hypertexnames=false,linkcolor=blue!60!black,
  citecolor=blue!60!black,urlcolor=blue!60!black,
  pdftitle={Ultraviolet dissipation, defect spectra, and transmission bounds in holographic interfaces},
  pdfauthor={Mianqi Wang}}

\newcommand{\dd}{\mathrm{d}}
\newcommand{\AdS}{\mathrm{AdS}}

\newcommand{\cUV}{c_{\mathrm{UV}}}
\newcommand{\ceff}{c_{\mathrm{eff}}}
\newcommand{\cLR}{c_{LR}}

\newcommand{\Amin}{A_{\min}}
\DeclareMathOperator{\Res}{Res}

\title{Ultraviolet dissipation and defect OPE in bottom-up holographic interfaces}

\author[a,b]{Mianqi Wang}

\affiliation[a]{Weinberg Institute for Theoretical Physics,
The University of Texas at Austin, Austin, TX 78712, USA}

\affiliation[b]{Kavli Institute for the Physics and Mathematics
of the Universe (WPI), The University of Tokyo Institutes for
Advanced Study, The University of Tokyo, Kashiwa, Chiba 277-8583, Japan}

\emailAdd{mianqi.wang@ipmu.jp}

\abstract{
We study the radial ultraviolet dissipation quantity $c_{UV}$ defined in \cite{karch2026universalitydissipationholographicinterfaces} and defect OPE data in bottom-up holographic interfaces. We relate $c_{UV}$ along with the defect primary spectrum and BOPE coefficients in the large defect dimension limit to the optical length $\rho$ in the holographic bulk. In addition, we prove a new holographic bound for conformal interfaces relating entanglement entropy, energy transmission and UV dissipation:
\[
c_{LR}\geq
\frac{2c_{\mathrm{eff}}}{2+\pi(c/c_{\mathrm{UV}}-1)}.
\]

This compliments the upper bound of $c_{LR}$ by $c_\text{eff}$ proved in \cite{Karch:2024udk}.}

\begin{document}
\maketitle

\section{Introduction}

Conformal interfaces in two-dimensional conformal field theories preserve a
diagonal conformal symmetry while allowing nontrivial reflection and
transmission of excitations.  Their equilibrium properties are described
by several universal observables.  Some of the well-studied quantities include
the effective central charge $\ceff$ that measures universal entanglement entropy across
the interface, and $\cLR$ that governs energy transmission
\cite{Karch:2022vot,Karch:2023evr,Karch:2024udk}. 

On the other hand, studying conformal interfaces in open systems coupled to an external environment provides a complementary measurement of how the interface suppresses dissipation. This approach was developed for critical free-fermion chains in
\cite{barad2025dissipationmeetsconformalinterface}, where a local Lindblad
term introduces dissipation on one side of the interface. The
ratio of Liouvillian gaps (lowest decay mode) with and without the interface defines another interface observable
$c_{\mathrm{relax}}$ that is universal in certain regimes. 

Holographic interfaces from bottom-up and top-down models are great playgrounds to study those universal observables. A much-studied family of top-down holographic interfaces are the Janus interfaces in deformation of AdS$_3\times S^3\times T^4$, dual to 2D $\mathcal{N}=(4,4)$ CFTs with non-BPS or half-BPS interfaces \cite{Bak_2007,Chiodaroli_2010}. More generally, one can consider bottom-up models of interface CFT (ICFT) that are described by a warpfactor function $A(x)$ on the AdS$_2$ slicing coordinates. It can help describe a large class of smooth bulk deformation of pure AdS$_3$, providing that $A(x)$ satisfies certain energy bounds and obeys Einstein's equation.

Solving fluctuations and normal modes of probe scalars in bottom-up and top-down models of holographic Janus ICFT was done in \cite{Chiodaroli:2016jod,gutperle2025januscorrelatorsheunsequation,karch2026universalitydissipationholographicinterfaces}. In particular, analytical solutions were given for all orders of jump parameters for BPS-Janus interfaces, but it was clear from \cite{gutperle2025januscorrelatorsheunsequation} that for generic Janus solutions we can only solve the spectrum perturbatively in the coupling jump. For generic bottom-up model, it would be more challenging to solve the Klein-Gordon equation even numerically.

In \cite{karch2026universalitydissipationholographicinterfaces}, a new interface quantity was defined for holographic interfaces, in particular for BPS-Janus. Similar to $c_\text{relax}$, it was an open-system quantity, after coupling one side of the interface via a double-trace deformation of scalars to a holographic thermal CFT dual to a BTZ black hole. But instead of considering the lowest decay rate of the entire system, a large-defect-dimension limit $n\to\infty$ was taken while fixing the AdS$_2$ descendant level $k$. In addition, one can consider the uniform coupling of half space instead of a local dissipation. The dissipation rate was extracted from the quasinormal modes of a probe scalar, and the ratio of the dissipation rates in that limit with and without the interface defines $c_{UV}$. Coincidentally, for BPS-Janus $c_{UV}/c=c_\text{eff}/c=1/\kappa$.

More recently, \cite{vargas2026holographicopenquantum} also considered the WKB limit where they considered the no-interface case and derived the full set of quasinormal modes analytically by studying the retarded Green's function. When conformal interfaces are present, even in the large-defect-dimension WKB limit the spectra of primaries mix with their descendant at the same frequency level and is very subtle to do perturbation theory\footnote{Already in the BPS-Janus case of \cite{karch2026universalitydissipationholographicinterfaces}, the degeneration of its normal poles depends heavily on the rationality of the jump parameter $\kappa$ and the level gaps at large frequency tends to zero.}. We instead focus on the radial WKB limit fixing the descendant level to avoid this subtlety.

In this paper, we show that taking this limit is very powerful. We are able to calculate $c_{UV}$ for a generic class\footnote{Subject to bound conditions on the warpfactor in Section \ref{sec:closed-modes}.} of bottom-up models. We further relate $c_{UV}$ to the bulk geometry via the warpfactor. Explicitly, for the bottom-up model
$\dd s^2=\dd x^2+A(x)^2\dd s^2_{\AdS_2}$, $c_{UV}$ is related to the optical length $\rho$ by
\begin{equation}
  \frac{c}{c_{UV}}=\rho=\frac{1}{\pi}\int_{-\infty}^{\infty}\frac{\dd x}{A(x)}.
  \label{eq:intro-optical-density}
\end{equation} 

This is in line with the calculations of closed-observables like $c_\text{eff}$ and $c_{LR}$ for holographic interfaces \cite{Karch_2021,BachasEtAl:2023ThickBranes}. In addition, from this geometric interpretation of $c_{UV}$ we prove a new lower bound of $c_{LR}$ in terms of $c_{\text{eff}}$ and $c_{UV}$ for any holographic conformal interface. This is in addition to its upper bound established in \cite{Karch:2024udk}. Concretely, for $c_1=c_2=c$ and warpfactor satisfying the null energy
condition (NEC), we prove $0\leq \cUV\leq c$, and
\begin{equation}
  0\le\frac{2\ceff}{2+\pi(c/\cUV-1)}\leq\cLR\leq\ceff\leq c.
  \label{eq:intro-charge-bounds}
\end{equation}

Another canonical set of ICFT data for probe operators is its BOPE or defect OPE (dOPE) data\footnote{We use dOPE and BOPE interchangeably in this paper.}, which involves its 1-pt and 2-pt functions in presence of the interface/defect. For the probe scalar in the bulk, its defect spectrum and BOPE coefficients are given by the fluctuation equations, so they are by definition closely related to the (quasi)normal modes and $c_{UV}$ \cite{Aharony_2003,Korovin:2014FirstOrder,
MelbyThompsonSchmidtColinet:2017DoubleTrace,
MazacRastelliZhou:2019AnalyticBCFT,BachasEstes:2011Spin2,
Chiodaroli:2016jod,gutperle2025januscorrelatorsheunsequation}. In this paper, we also sort out this relation and observe that in this large-defect-dimension limit, the ratio of BOPE coefficients with and without the interface is also directly given by $1/\rho=c_{UV}/c$.

The paper is organized as follows: In Section~\ref{sec:closed-modes} we develop the closed spectrum and BOPE data of the bottom-up holographic model.
In Sections~\ref{sec:uniform-btz} and \ref{sec:poles} we construct the projected
system--bath response in the large-defect-dimension limit and derive the dissipation rate in the double-trace equations and subsequently $c_{UV}$. In Section \ref{sec:closed-bope} we establish the relation between the BOPE coefficients, defect spectrum, $c_{UV}$ and bulk geometry.
Section~\ref{sec:new-bound} proves the new holographic bounds and their saturation
conditions.  The appendices supplements the proof of the bound and provides checks of our results on pure-$\AdS_3$ and known Janus examples.

\section{Normal modes of a general bottom-up holographic interface}
\label{sec:closed-modes}

In this section we consider the bottom-up holographic ICFT and study the normal modes of probe scalar at fixed descendant level at large defect dimension. We will see that the density of defect states $\rho$ in this limit has a natural holographic interpretation in terms of bulk geometry. Furthermore, we establish that in this limit, $\rho$ is also related to the BOPE coefficients of the same holographic interface.

Consider the standard bottom-up holographic bulk of a 2D interface CFT \cite{Karch_2021}. 
The metric is
\begin{equation}
  \dd s^2=\dd x^2+A(x)^2\dd s^2_{\AdS_2}.
  \label{eq:closed-metric}
\end{equation}
 with $x\in(-\infty,\infty)$ transverse to the interface and warpfactor $A(x)$. Here the unit global-$\AdS_2$ slice is
$\dd s^2_{\AdS_2}=\sec^2\varphi(-\dd t^2+\dd\varphi^2)$, with global time
$t$ and $\varphi\in(-\pi/2,\pi/2)$. The interfaces are located at $\varphi=\pm\pi/2$, extended along the time direction. We set a common unit asymptotic
$\AdS_3$ radius near the boundaries, so that $A(x)\sim a_-e^{-x}$ as $x\to-\infty$ and
$A(x)\sim a_+e^x$ as $x\to+\infty$, where $a_\pm>0$.

Let us now consider a probe scalar in this geometry. The probe action is
\begin{equation}
  S_\Phi=-\frac12\int\dd t\,\dd x\,\dd\varphi\,\sqrt{-g}\,
  \left[(\partial\Phi)^2+m_\Phi^2\Phi^2\right].
  \label{eq:scalar-action}
\end{equation}

\subsection{The Klein-Gordon equation}
\label{sec:closed-separation}

Let us solve the Klein-Gordon (K-G) equation in this geometry. Separation of variables gives us the eigenvalue problem in terms of AdS$_2$ modes. This will determine the defect-primary dimensions $d_n$
and the $\AdS_2$ equation which fixes their descendant
levels.

Write the standard-quantization dimension as $\Delta=1+\nu$ with $\nu>0$,
so the bulk mass is $m_\Phi^2=\nu^2-1=\Delta(\Delta-2)$.  We use standard
quantization at both asymptotic boundaries.  The AdS$_2$ dimension is
$d>1/2$, and mass $q(d)^2=d(d-1)$.

The solution of the K-G equation is
 \begin{equation}
   \Phi=e^{-i\omega t}f(x;d)\chi(\varphi;\omega,d),
   \label{eq:closed-separated-mode}
 \end{equation}
where the transverse function obeys
 \begin{equation}
   \partial_x(A^2\partial_xf)
   +\left[q(d)^2-m_\Phi^2A^2\right]f=0,
   \label{eq:closed-radial-equation}
 \end{equation}
and the global-$\AdS_2$ factor obeys
 \begin{equation}
   \partial^2_\varphi\,\chi
   +\left[\omega^2-q(d)^2\sec^2\varphi\right]\chi=0.
   \label{eq:closed-ads2-equation}
 \end{equation}
 Imposing the source-free standard-quantization behavior at both boundaries
$x\to\pm\infty$ quantizes the eigenvalues of \eqref{eq:closed-radial-equation} to be $q_n^2$,
$n=0,1,\ldots$. The AdS$_2$ dimensions are $d_n=1/2+\sqrt{1/4+q_n^2}$.

Imposing the standard-quantization global-$\AdS_2$
boundary conditions at $\varphi=\pm\pi/2$ quantizes the frequency $\omega$. There are two parity sectors on $\varphi\in(-\pi/2,\pi/2)$
with energies $d_n+2p$ and $d_n+2p+1$, with $p\in\mathbb N_0$. Combining
them into one descendant label $k\in\mathbb N_0$ gives the normal modes
\begin{equation}
  \omega_{n,k}^{(\pm)}=\pm(d_n+k).
  \label{eq:closed-exact-normal-modes}
\end{equation}
We denote the positive frequency by $\Omega_{n,k}=d_n+k$.
The properly normalized eigenfunction is from \cite{karch2026universalitydissipationholographicinterfaces}: for $\lambda_k \equiv \omega-k$ with $k=0,1,2,\ldots$,
\begin{align}
  \chi_k(\varphi;\omega)
  &=
  \frac{2^{\lambda_k-1}\Gamma(\lambda_k)}{\sqrt{\pi}}
  \left[\frac{\Gamma(k+1)}{\Gamma(k+2\lambda_k)}\right]^{1/2}
  (\cos\varphi)^{\lambda_k}
  C_k^{(\lambda_k)}(\sin\varphi).
  \label{eq:ads2eigen}
\end{align}

Here $C_k^{(d_n)}$ is the Gegenbauer polynomial of degree $k$ and index
$d_n$.
Equivalently, the poles of the closed cylinder two-point function occur at
$\omega=\pm(d_n+k)$, which is the standard state--operator interpretation of
the defect BOPE tower \cite{Aharony_2003,BilloEtAl:2016Defects}. Here we have fixed its normalization by
imposing the bilinear normalization
\begin{equation}
  \int_{-\pi/2}^{\pi/2}\dd\varphi\,
  \sec^2\varphi\,\chi_k(\varphi;\omega)^2
  =\frac1{2\lambda_k-1}.
  \label{eq:channel-normalization}
\end{equation}
At $\omega=\Omega_{n,k}$ this is the real on-shell normalization
$\int\sec^2\varphi\,\chi_{n,k}^2\,\dd\varphi=1/(2d_n-1)$.

\subsection{Defect spectra in the WKB limit}
\label{sec:closed-optical}

For a generic warpfactor $A(x)$ the above Sturm-Liouville eigenvalue problem is very hard to solve. However, we can still use its asymptotic behavior to extract information about the normal modes, in particular, the density of states at large-$n$ fixing $k$. We consider the optical coordinate of the bulk geometry which
converts the radial equation into a finite-interval Schr\"odinger equation. In this sense the large-$n$ fixed-$k$ limit is precisely the WKB limit, and we can relate the density of large radial defect states to the geometric optical coordinate.

Define the optical coordinate as the integral
\begin{equation}
  u(x)=\int_{-\infty}^{x}\frac{\dd x'}{A(x')}.
  \label{eq:closed-optical-data}
\end{equation}
Its range is $u\in(0,L)$, where the optical length is
\begin{equation}
  L=\int_{-\infty}^{\infty}\frac{\dd x}{A(x)}.
  \label{eq:closed-optical-length}
\end{equation}
We normalize the asymptotic radial density as $\rho=L/\pi$. Rescale $f=A^{-1/2}\Psi$, the radial
equation \eqref{eq:closed-radial-equation} is then
\begin{equation}
  -\Psi''+V_{\Delta}(u)\Psi=q^2\Psi,
  \label{eq:closed-optical-Schrodinger}
 \end{equation}
where the Liouville potential is
\begin{equation*}
  V_{\Delta}
  =m_\Phi^2A^2+\frac{\partial_u^2A}{2A}
   -\frac14\left(\frac{\partial_uA}{A}\right)^2.
\end{equation*}

  Recall from the near-boundary asymptotics of $A$, near $x\to-\infty$ we have $u\sim e^x/a_-$. This gives $A(u\to 0)=u^{-1}[1+o(1)]$. For the derivations below, we further make the assumption on the asymptotic behavior of the warpfactor and its derivatives: for $y=u\to 0$ on the left side and $y=L-u\to 0$ on the right side, we demand
\begin{equation}
  A(y)=y^{-1}[1+\varepsilon(y)]
\end{equation}  
  
where the function $\varepsilon$ satisfies
\begin{equation}    
\varepsilon(y)=O(y^2),\quad \partial_y\,\varepsilon(y)=O(y),\quad
  \partial_y^2\,\varepsilon(y)=O(1).
  \label{eq:assumption1}
\end{equation}

One can check that most holographic ICFT in 3D satisfies this stronger assumption\footnote{3D Janus \cite{Bak_2007}, BPS-Janus \cite{Chiodaroli_2010} and standard $\mathcal{N}=8$ gauged supergravity Janus \cite{Chen_2021} satisfies this, while its certain deformation in \cite{Chen_2022} and $\mathcal{N}=2$ gauged supergravity Janus in \cite{gutperle2024janusrginterfacesminimal3d} fails this assumption. See Appendix \ref{app:janus-examples} for details.}.  From the above conditions, near the left boundary the Liouville potential gives
\begin{equation}
  V_{\Delta}(u)
  =\frac{(\Delta-1)^2-\frac14}{u^2}+O(1),
  \label{eq:closed-Bessel-endpoint}
\end{equation}
and similarly for $L-u$.

To solve the eigenfunction, let us now introduce a reference potential $V_0$ to approximate the full $V_\Delta=V_0(u)+W_\nu(u)$. From \eqref{eq:assumption1}, we have
\begin{equation}
  V_0(u)=\left(\nu^2-\frac14\right)
    \left(\frac{\pi}{L}\right)^2
    \csc^2\left(\frac{\pi u}{L}\right),
  \label{eq:bounded-endpoint-remainder}
\end{equation}
and $W_\nu=V_\Delta-V_0$ is a bounded function in the domain $(0,L)$, meaning that
its essential supremum $\|W_\nu\|_\infty$ is finite for any $q$.
The reference potential matches with $V_\Delta$ near the two boundaries.

After rescaling, the reference operator
$(-\partial_u^2+V_0)$ on $\Psi$ has exact eigenvalues for $f$
\begin{equation*}
  E_n^{(0)}=\left(\frac{\pi}{L}\right)^2
                \left(n+\nu+\frac12\right)^2,
  \qquad n=0,1,\ldots.
\end{equation*}
The min--max principle then demands that each eigenvalue of $q^2$ is close to the eigenvalue of $E_n^{(0)}$
\begin{equation}
  \left|q_n^2-E_n^{(0)}\right|\leq\|W_\nu\|_\infty.
  \label{eq:bounded-remainder-eigenvalues}
\end{equation}
Hence at large $n$,
\begin{equation}
  q_nL=\pi\left(n+\nu+\frac12\right)+O(n^{-1}).
  \label{eq:closed-WKB-q}
\end{equation}
Plugging in $d_n=\frac12+\sqrt{\frac14+q_n^2}$ and the normal modes $\Omega_{n,k}=d_n+k$ give, at fixed $k$,
\begin{equation}
  \Omega_{n,k}
    =k+\frac12+
       \frac{n+\Delta-\frac12}{\rho}+O(n^{-1}),
  \label{eq:closed-general-WKB-spectrum}
\end{equation}
where recall that $\rho= L/\pi$. Define the staircase counting function by
$N_k(\omega)=\#\{n:\Omega_{n,k}\leq\omega\}$.
Its Weyl asymptotic is
\begin{equation}
  N_k(\omega)=\frac L\pi
    \sqrt{(\omega-k)(\omega-k-1)}+O(1),
  \label{eq:closed-counting-law}
\end{equation}
and hence
\begin{equation}
  \lim_{\omega\to\infty}\frac{N_k(\omega)}{\omega}=\rho.
  \label{eq:closed-density}
\end{equation}

Thus the optical length quantity $\rho$ is the asymptotic density of the defect-primary tower at large AdS$_2$ dimension $d$, at any fixed AdS$_2$ momenta level $k$. In the language of defect operator product expansion (dOPE) \cite{Aharony_2003,BilloEtAl:2016Defects,LauriaMeineriTrevisani:2018RadialDefect}, $d_n$ are the dimensions for defect primaries of the probe scalar operator, and $k$ is the descendant level \cite{wang2026twistoperatorbopeentanglement}. We will use these names interchangeably in this paper\footnote{Note that $\rho$ is not the density of defects states at a given defect dimension $\omega$, since  $(n,k)$ with large $k$ will also mix together. Here we only consider the radial limit fixing $k=O(1)$.}.

\section{Uniform coupling to a BTZ bath}
\label{sec:uniform-btz}

In the last section we established that in presence of a conformal interface that has a holographic dual, with some minor restrictions on the bulk geometry, the density of defect primaries of a probe scalar at large dimension (or equivalently, the density of normal modes) is related to the optical length $\rho$ of the asymptotic AdS dual. Below we will turn to another interface quantity that was related to quasinormal modes in a open holographic system: $c_{UV}$ in \cite{karch2026universalitydissipationholographicinterfaces}. As we will review below, it measures how the holographic interface suppresses the dissipation rate once the system CFT is coupled to a thermal bath on one side of the interface. In \cite{karch2026universalitydissipationholographicinterfaces}, $c_{UV}$ was derived for BPS-Janus. Interestingly, we will show that this open system quantity is also related to the same $\rho$ in the bottom-up ICFT. 

Following \cite{karch2026universalitydissipationholographicinterfaces} we setup the system and bath coupled via a uniform double-trace (DT) deformation of scalars over the entire half space. Let
$\mathcal H=(-\pi/2,\pi/2)$ denote the angular half-circle on the boundary
that will be uniformly coupled to the bath, so that
\begin{equation}
  S_{\mathrm{int}}
  =-h_R\int_{\mathcal{H}}\dd t\,\dd\varphi\,
    X_B\mathcal O_S.
  \label{eq:interaction}
\end{equation}

We generalize the massless scalar with marginal double-trace coupling there, and consider a generic massive scalar in the bottom-up holographic model at hand with marginal couplings.

Let us couple the right half-space of the zero-T interface theory (system) to a
second thermal CFT dual to a BTZ black hole (bath). On the left half-space they are completely disentangled.  

Throughout the paper we still keep the interaction a uniform (DT) deformation supported on the
coupled half-cylinder as in \cite{karch2026universalitydissipationholographicinterfaces}. In the holographic description, it relates the
source and response coefficients of scalar fields in the two asymptotically
AdS halves through reciprocal mixed boundary conditions
\cite{Witten:2001ua,Aharony:2006hz,Kiritsis:2006hy,
karch2026universalitydissipationholographicinterfaces}. 

The massless case in
\cite{karch2026universalitydissipationholographicinterfaces} has
$\nu=1$ and additional subtleties in terms of the Fefferman-Graham expansion, and here instead we take the generic scalar
$m_S^2\equiv m_\Phi^2$ within the unitary bound
\begin{equation}
  m_S^2=m_B^2=\nu^2-1,
  \qquad 0<\nu<\frac{1}{2}.
  \label{eq:matched-range}
\end{equation}
Standard system quantization gives $\Delta_S=1+\nu$, while the bath field is in alternate quantization with
$\Delta_{B,-}=1-\nu$.  Both quantizations are unitary in this window
\cite{BreitenlohnerFreedman:1982PositiveEnergy,
Klebanov:1999tb,Witten:2001ua}. 

For the current system, like in \cite{karch2026universalitydissipationholographicinterfaces}, the geometry is AdS$_2$ slicing of asymptotic AdS$_3$, so we need to be careful matching the response and source of system with the bath in the double-trace equations. We first derive the system and bath responses and their large dimension limit, before writing down the full DT equations in the descendant-level matrix form. We leave the numerical quasinormal solutions of the full system+bath to future work.  Later in Section \ref{sec:poles} we will project the problem to the large defect dimension limit with fixed descendant level $k$ and analytically solve the DT equations up to leading order.

\subsection{System response}
\label{sec:closed-response}

We begin with the system bottom-up holographic ICFT by first calculating the scalar response Green's functions. For a generic warp factor it is difficult to obtain analytical results, but by taking the UV $n\to\infty$ limit while keeping $k=O(1)$ we can fix the leading orders by comparing with the near-boundary pure AdS$_3$ expressions.

We use the optical coordinate $u$ in \eqref{eq:closed-optical-data} of the system geometry throughout.
Let $y$ be inward distance from the chosen boundary: $y=u$ at the left
face and $y=L-u$ at the right side. The metric is
\begin{equation*}
  \dd s_S^2=A(u)^2\left[
    \dd u^2+\sec^2\varphi(-\dd t^2+\dd\varphi^2)\right].
\end{equation*}
At fixed descendant level $k$, set
$\lambda_k=\omega-k$ and $E_k=\lambda_k(\lambda_k-1)$.  From the solution in \eqref{eq:closed-separated-mode} we can write down the solution channel by channel in terms of the AdS$_2$ Gegenbauer label $k$:
$\Phi_k^S=e^{-i\omega t}f_k(u;\omega)\chi_k(\varphi;\omega)$.

where $f_k$ is the solution to
\begin{equation}
  \partial_u\!\left(A(u)\partial_u f_k\right)
  +A(u)\left[E_k-(\nu^2-1)A(u)^2\right]f_k=0,
  \label{eq:matrix-bottom-up-radial}
\end{equation}
and $\chi_k$ is the AdS$_2$ eigenfunction in \eqref{eq:ads2eigen}.

For the noninteger dimension $\nu$, near either boundary the radial source and response amplitudes are
\begin{equation}
  f_k(y;\omega)
  =\alpha_k^S(\omega)y^{1-\nu}
   +\beta_k^S(\omega)y^{1+\nu}+\cdots .
  \label{eq:closed-massive-FG}
\end{equation}

Here $\alpha_k^S(\omega)
   ,\,\beta_k^S(\omega)$ are the radial source and response at fixed $(\omega,k)$, with their ratio fixed
by the source-free condition at the opposite face.

For later purposes where we couple this AdS$_2$ slicing geometry to BTZ in cylinder coordinates, it is good to sort out the matching of coordinate basis. In the system metric we have the boundary metric
$\sec^2\varphi(-\dd t^2+\dd\varphi^2)$.  The coupling to BTZ is specified
instead in the cylinder form $-\dd t^2+\dd\varphi^2$,
obtained by a boundary Weyl rescaling by $\cos^2\varphi$.
For $\Delta=1+\nu$, the source has the rescaling weight $\Delta-2$ and the
response has weight $-\Delta$. We suppress the explicit $\omega$ argument of $\chi_k$ from now on and define
the source and response profiles by
\begin{align}
  s_{\nu,k}(\varphi)
  &=(\cos\varphi)^{\nu-1}\chi_k(\varphi), \nonumber\\
  v_{\nu,k}(\varphi)
  &=(\cos\varphi)^{-\nu-1}\chi_k(\varphi).
  \label{eq:source-response-profiles}
\end{align}
They obey the exact cross-normalization
\begin{equation}
  (2\lambda_k-1)
  \int_{-\pi/2}^{\pi/2}\dd\varphi\,
  s_{\nu,k}(\varphi)v_{\nu,k}(\varphi)=1.
  \label{eq:cross-normalization}
\end{equation}

Then one can denote the cylinder source and response
profiles of the individual channel by $a_k^S$ and $b_{k,R}^S$:
\begin{equation}
  a_k^S(\omega,\varphi)
    =\alpha_k^S(\omega)s_{\nu,k}, \qquad
  b_{k,R}^S(\omega,\varphi)
    =\beta_k^S(\omega)v_{\nu,k}
       +\text{local source terms}.
  \label{eq:closed-massive-boundary-profiles}
\end{equation}

For the most general sources we have the superposition of channels for the sources and responses: 
$a^S=\sum_j C_j^S a_j^S$ and $b_R^S=\sum_j C_j^S b_{j,R}^S$. The renormalized
on-shell action varies as
\begin{equation}
  \delta S_S^{\mathrm{ren}}
  =\int_{\mathcal H}\dd t\,\dd\varphi\,
    2\nu b_R^S\,\delta a^S.
\end{equation}  
 At fixed frequency, the full same-face retarded response is the integral
kernel defined by
\begin{equation}
  2\nu b_R^S(\omega,\varphi)
  =(G_R^S(\omega)a^S)(\varphi)
  =\int_{\mathcal H}\dd\varphi'\,
     G_R^S(\omega;\varphi,\varphi')a^S(\omega,\varphi').
  \label{eq:full-system-response-kernel}
\end{equation}

In Section~\ref{sec:double-trace} we will write these sums out and constructs
the coupled mode matrix for solving the exact quasinormal modes after coupling to the bath. For now (and for the purpose of the large-radial-dimension projection) we only consider a source $a^S=\alpha_k^S s_{\nu,k}$ purely in the $k$th channel, and the response is
\begin{equation}
  G_R^S(\omega)s_{\nu,k}
  =\frac{2\nu\beta_k^S}{\alpha_k^S}v_{\nu,k}
    +\mathcal L_\nu(\omega)s_{\nu,k},
  \label{eq:full-response-on-channel}
\end{equation}
where $\mathcal L_\nu$ is the local contact operator in the chosen
renormalization prescription when there are local terms in the on-shell action \cite{Bianchi:2001kw}. For the noninteger dimension at hand we can choose the scheme $\mathcal{L}_\nu=0$, but it is important for the discussion of massless scalars like the ones in \cite{karch2026universalitydissipationholographicinterfaces}. 

Using the analytic bilinear pairing
$\langle f,g\rangle=\int_{\mathcal H}f(\varphi)g(\varphi)\,\dd\varphi$, we can extract a 'projected response function' $G_{R,k}^{S,\mathrm{proj},(\nu)}$ on channel $k$ that is given by the ratio of $\beta_k$ and $\alpha_k$:
\begin{align}
  G_{R,k}^{S,\mathrm{proj},(\nu)}(\omega)
  &\equiv(2\lambda_k-1)
       \langle s_{\nu,k},G_R^S(\omega)s_{\nu,k}\rangle
       \nonumber\\
  &=(2\lambda_k-1)\int_{\mathcal H}\dd\varphi
       \int_{\mathcal H}\dd\varphi'\,
       s_{\nu,k}(\varphi)G_R^S(\omega;\varphi,\varphi')
       s_{\nu,k}(\varphi') .
  \label{eq:closed-massive-projected-response}
\end{align}
Equivalently, using \eqref{eq:full-system-response-kernel}, this is
$2\nu(2\lambda_k-1)(\alpha_k^S)^{-1}
\int_{\mathcal H}s_{\nu,k}b_R^S\,\dd\varphi$.
Substituting \eqref{eq:full-response-on-channel} and
\eqref{eq:cross-normalization} gives
\begin{equation}
  G_{R,k}^{S,\mathrm{proj},(\nu)}(\omega)
  =\frac{2\nu\beta_k^S(\omega)}{\alpha_k^S(\omega)}
     +G_{\mathrm{loc}}^{(\nu)}(\omega), 
  \label{eq:projected-radial-ratio}
\end{equation}
 where $G_{\mathrm{loc}}^{(\nu)}
  =(2\lambda_k-1)\langle s_{\nu,k},\mathcal L_\nu s_{\nu,k}\rangle$ is the local scheme-dependent counterterm.
  
Throughout the paper the label $k$ specifies the AdS$_2$ channel and $n$ labels the radial normal modes
$\Omega_{n,k}=d_n+k$. 
Set $v_{n,k}=v_{\nu,k}(\,\cdot\,;\Omega_{n,k})$ and write the residue
$\mathcal R_{n,k}^S=-\Res_{\omega=\Omega_{n,k}}
G_{R,k}^{S,\mathrm{proj},(\nu)}$ at the pole $(n,k)$,
the contribution of this normal mode to the full $G_R^S$ is
\begin{equation}
  -\frac{(2d_n-1)\mathcal R_{n,k}^S}
         {\omega-\Omega_{n,k}+i0}\,
       v_{n,k}(\varphi)v_{n,k}(\varphi').
  \label{eq:full-kernel-pole-contribution}
\end{equation}

For a generic warpfactor $A(u)$,
$\beta_k^S/\alpha_k^S$ and $\mathcal{R}_{n,k}^S$ are not available in closed form and must be evaluated numerically at finite frequency. In the following sections we derive their leading expressions by first taking the large defect dimension limit $n\to\infty$ and then take the near-boundary limit where their expressions are known for asymptotic AdS$_3$.

We now take $n\to\infty$ at fixed descendant level $k$.  Near the fixed frequency $\omega=\Omega_{n,k}+O(1)$, recall $q=\sqrt{\lambda_k(\lambda_k-1)}\sim\frac{\pi n}{L}\to\infty$ where $\lambda_k=\omega-k$. Let us focus on the left side near-boundary expansion, and the ratio $\beta_k^S/\alpha_k^S$ is fixed by the solution normalizable
near the right boundary.  

In the following we first define the endpoint solutions
and propagation phase exactly for the solution to \eqref{eq:closed-optical-Schrodinger}, then obtain their large-$q$ estimates
by matching the Bessel functions asymptotically. Recall that we used a reference potential for the entire $u\in[0,L]$ to obtain the spectra at large dimension, but here we use the weaker version of \eqref{eq:closed-Bessel-endpoint}. By first taking the UV limit we approximate the exact solution with the reference solution in almost the entire range of $u$.

Let $\Psi_L(u;q)$ and $\Psi_R(u;q)$ denote the exact
solutions of \eqref{eq:closed-optical-Schrodinger} normalized by the boundary conditions 
\begin{equation}
  \lim_{u\to0}u^{-\nu-1/2}\Psi_L(u;q)=1,\qquad
  \lim_{u\to L}(L-u)^{-\nu-1/2}\Psi_R(u;q)=1.
  \label{eq:exact-solution}
\end{equation}
These conditions select
the standard-quantization solution
$\Psi\propto y^{\nu+1/2}$ for $y=u$ or $L-u$, or equivalently
$f=A^{-1/2}\Psi\propto y^\Delta$.

Using the approximation of the potential in \eqref{eq:closed-Bessel-endpoint} which matches on the left boundary, we solve the  comparison
solution $\Psi$ to be
\begin{equation}
  \Psi_\text{comp}(u)
  =\Gamma(\nu+1)\left(\frac2q\right)^\nu
       \sqrt{u}\,J_\nu(qu).
       \label{eq:cmpsolution}
\end{equation}
Here $J_\nu$ is the Bessel function of the first kind. In the large $q$ limit (by first taking $q\to\infty$) we then have the leading order expression of the exact solution $\Psi_L$, $\Psi_R$ for any range
$[\epsilon,L-\epsilon]$ with arbitrary positive number $0<\epsilon<L/2$ \cite{KostenkoSakhnovichTeschl:2010Inverse,KostenkoTeschl:2011SingularWeyl}:
\begin{align}
  \frac{\Psi_L(u;q)}{C_\nu(q)}
    &=\cos(qu-\alpha_\nu)+O(q^{-1}),&
  \frac{\Psi_L'(u;q)}{qC_\nu(q)}
    &=-\sin(qu-\alpha_\nu)+O(q^{-1}),\nonumber\\
  \frac{\Psi_R(u;q)}{C_\nu(q)}
    &=\cos(q(L-u)-\alpha_\nu)+O(q^{-1}),&
  \frac{\Psi_R'(u;q)}{qC_\nu(q)}
    &=\sin(q(L-u)-\alpha_\nu)+O(q^{-1}),
  \label{eq:closed-interior-solutions}
\end{align}

Here $\alpha_\nu=\pi\nu/2+\pi/4$ and $C_\nu(q)=
  2^\nu\Gamma(\nu+1)\sqrt{\frac2\pi}\,q^{-\nu-1/2}$.

Let $\Psi_{\mathrm{src}}(u;q)$ be
the exact left source solution, with unit coefficient of
$u^{1/2-\nu}$ and zero independent response coefficient in the chosen
boundary convention. Together with $\Psi_L$, it forms a basis of
solutions of the full radial equation. Thus
\begin{equation}
      \Psi_R(u;q)=\alpha_k^S\Psi_{\mathrm{src}}(u;q)
             +\beta_k^S\Psi_L(u;q).
             \label{eq:leftrightexact}
\end{equation}

Since the original field $f=A^{-1/2}\Psi$ and $A(u)\sim u^{-1}$, these are precisely
the coefficients of the source and response branches of the original
field, $u^{1-\nu}$ and $u^{1+\nu}$.

Define $\mathcal W[F,G]=FG'-F'G$. From the radial equation
gives $\partial_u\mathcal W=0$ and the boundary condition
$\mathcal W[\Psi_{\mathrm{src}},\Psi_L]=2\nu$, we have
\begin{equation}
      \alpha_k^S=-\frac{\mathcal W[\Psi_L,\Psi_R]}{2\nu},
  \qquad
  \beta_k^S=\frac{\mathcal W[\Psi_{\mathrm{src}},\Psi_R]}{2\nu}.
\end{equation}

 In particular,
$\mathcal W[\Psi_L,\Psi_R]=0$ is the condition that the
right-normalizable solution also has zero source at the left boundary, which are precisely the normal modes.

It is useful to express the boundary condition in terms of an
oscillatory phase and recover the defect spectra in this limit. Let us choose a fixed interior point
$u_*\in(0,L)$ and define real amplitudes $r_L,r_R>0$ and continuous
angles $\theta_L(q),\theta_R(q)$ by
\begin{equation}
  \begin{aligned}
    \Psi_L(u_*;q)&=r_L\cos\theta_L(q),
    &\Psi_L'(u_*;q)&=-q r_L\sin\theta_L(q),\\
    \Psi_R(u_*;q)&=r_R\cos\theta_R(q),
    &\Psi_R'(u_*;q)&= q r_R\sin\theta_R(q).
  \end{aligned}
\end{equation}

Choose the angle branches to agree with
\eqref{eq:closed-interior-solutions} at large $q$, and set
\begin{equation}
  \Theta_\nu(d)=\theta_L(q(d))+\theta_R(q(d)),
  \label{eq:closed-phase-definition}
\end{equation}
where $q(d)=\sqrt{d(d-1)}>0$. Substitution into the Wronskian gives
\begin{equation}
     \mathcal W[\Psi_L,\Psi_R]
     =q r_Lr_R\sin\Theta_\nu(d), 
\end{equation}

and the source
\begin{equation}
      \alpha_k^S=-\frac{q r_Lr_R}{2\nu}
      \sin\Theta_\nu(\omega-k).
\end{equation}

We can see that the normal modes are given by
$\Theta_\nu(d_n)=n\pi$. At fixed $k$, denote
$\Theta_{k}(\omega)\equiv\Theta_\nu(\omega-k)$. From \eqref{eq:closed-interior-solutions} we have
$\theta_L=qu_*-\alpha_\nu+O(q^{-1})$ and
$\theta_R=q(L-u_*)-\alpha_\nu+O(q^{-1})$, so
\begin{equation}
  \Theta_{\nu,k}(\omega)
  =q(d)L-\pi\left(\nu+\frac12\right)+O(q^{-1}).
  \label{eq:closed-WKB-phase}
\end{equation}

We now go back to the original solution $f=A^{-1/2}\Psi$ and determine the relative source and response amplitudes by
connecting the interior oscillation above to the left side near boundary expansion. Near the boundary $y=u,L-u$, the standard comparison solutions with unit F-G coefficients for the original field are respectively, $\mathsf S_{q}(y)
    =\Gamma(1-\nu)\left(\frac q2\right)^\nu
      yJ_{-\nu}(qy)$ and $\mathsf N_{q}(u)=\Gamma(1+\nu)\left(\frac 2q\right)^\nu
      yJ_{\nu}(qy)$. From the above analysis, after taking the large $q$ limit and make the large $qu$ approximation valid for $u\in[\epsilon, L-\epsilon]$, the leading order of $f$ is then
\begin{equation}
    f(y)
  =\mathcal A(q)y\left[
     J_\nu(qy)\cos\Theta_{k}(\omega)
    +Y_\nu(qy)\sin\Theta_{k}(\omega)\right].
\end{equation}

 Using the connection identity
$ Y_\nu=\cot(\pi\nu)J_\nu-\csc(\pi\nu)J_{-\nu}$, the F-G expansion of the leading order solution is
$f_\mathrm{comp}=
 \alpha_{k}^S\mathsf S_{q}
 +\beta_{k}^S\mathsf N_{q}$ where
\begin{equation}
  \begin{aligned}
    \alpha_{k}^S
      &=-\frac{\mathcal A(q)\csc(\pi\nu)
           \sin\Theta_{k}(\omega)}
          {\Gamma(1-\nu)(q/2)^\nu},\\
    \beta_{k}^S
      &=\frac{\mathcal A(q)
          [\cos\Theta_{k}(\omega)
           +\cot(\pi\nu)\sin\Theta_{k}(\omega)]}
          {\Gamma(1+\nu)(2/q)^\nu}.
  \end{aligned}
\end{equation}

 We note that any subleading term in $q$ has been suppressed. Using
$\Gamma(1+\nu)=\nu\Gamma(\nu)$ and
$\Gamma(\nu)\Gamma(1-\nu)=\pi/\sin(\pi\nu)$, we obtain the response function of the system in this projected limit
\begin{equation}
  G_{R,k}^{S,\mathrm{proj},(\nu)}(\omega)=2\nu\frac{\beta_{k}^S}
                {\alpha_{k}^S}
  =-\frac{2\pi}{\Gamma(\nu)^2}
     \left(\frac q2\right)^{2\nu}
     \left[\cot\Theta_{\nu,k}(\omega)+\cot(\pi\nu)\right].
     \label{eq:system-response-leading}
\end{equation}
 It is similar to the results in \cite{karch2026universalitydissipationholographicinterfaces} apart from the logarithm coefficients from the massless scalars.

Note that this leading order expression is valid away from the normal modes, but differs near the poles where the large $\cot\Theta$ term can shift the regular terms. We have also ignored the local contact term $G_\text{loc}$ that would be important for the massless case\footnote{When $\nu=N\in\mathbb N$, as in \cite{karch2026universalitydissipationholographicinterfaces},
 the two Bessel branches resonate and the FG
expansion contains a local $y^{N+1}\log y$ branch.  Its subtraction
introduces the system renormalization scale $\mu_S$
\cite{Bianchi:2001kw,Skenderis:2002wp}. The local term $G_\text{loc}$ now matters since we need to carefully deal with the contact terms. Continuing the
noninteger response through $\nu=N$ and subtracting the local divergence gives
\begin{align}
  G_{R,k}^{S,\mathrm{proj},(N)}(\omega)
  ={}&-\frac{2}{\Gamma(N)^2}
  \left(\frac q2\right)^{2N}
  \left[
    \pi\cot\Theta_{N,k}(\omega)
    +2\log\left(\frac{q}{2\mu_S}\right)
  \right]
  \nonumber\\
  &+\frac{2}{\Gamma(N)^2}
  \left(\frac q2\right)^{2N}
  [\psi_0(1)+\psi_0(N+1)]
  +P_N(q^2)+o(q^{2N}),
  \label{eq:closed-massive-integer-response}
\end{align}
where $P_N$ is a real scheme-dependent local polynomial of degree at most
$N$ in $q^2$ and $\psi_0$ is the digamma function.}. In the next section we will calculate the residue of the response function $\mathcal{R}_{n,k}^S$ at each normal mode pole to compare with the BOPE coefficients.

\subsection{BTZ bath response}
\label{sec:btz-projection}

Let us now turn to the bath. Take the bath to be nonrotating BTZ, with radial coordinate $r$, horizon
radius $r_+$, and boundary angle $\varphi\sim\varphi+2\pi$:
\begin{equation}
  \dd s_B^2
  =-(r^2-r_+^2)\dd t^2
   +\frac{\dd r^2}{r^2-r_+^2}
   +r^2\dd\varphi^2.
  \label{eq:btz-metric}
\end{equation}
Its Hawking temperature is $T=r_+/(2\pi)$. The mass of the scalar operator $X_B$ is taken to be the same as $m_S$. Its near-boundary expansion is
\begin{equation}
  \Phi^B
  =r^{\nu_B-1}[a^B+\cdots]
   +r^{-\nu_B-1}[b_R^B+\cdots].
  \label{eq:bath-fg}
\end{equation}
Here $a^B$ and $b_R^B$ are its two renormalized
F-G coefficients. We use the alternative convention that
$\langle X_B\rangle=G_{R,-}^B b_R^B$ and
$a^B=-(2\nu_B)^{-1}G_{R,-}^B b_R^B$.

For each Fourier mode
$e_m(\varphi)=e^{im\varphi}/\sqrt{2\pi}$ over $\phi\in(-\pi,\pi]$, choose a fixed normalization
of the ingoing radial solution $\mathcal R_m^B$. Then we can consider the F-G expansion mode by mode from
\begin{equation}
  \Phi_m^B
  =r^{\nu_B-1}[a_m^B(\omega)+\cdots]
   +r^{-\nu_B-1}[b_{m,R}^B(\omega)+\cdots].
  \label{eq:bath-mode-coefficients}
\end{equation}

 The response function in these diagonal cylinder modes are then
$$G_{R,-}^B(\omega,m)=-2\nu_B a_m^B/b_{m,R}^B.$$

For the Fourier mode $e^{-i\omega t+im\varphi}$, $m\in\mathbb Z$,
the standard results from ingoing hypergeometric solution gives
\cite{karch2025dissipationopenholography,SonStarinets:2002Retarded,
BirminghamSachsSolodukhin:2002BTZ,KeskiVakkuri:1998BTZ}
\begin{align}
  G_{R,-}^B(\omega,m)
  ={}&-2\nu_B r_+^{-2\nu_B}
  \frac{\Gamma(\nu_B)}{\Gamma(-\nu_B)}
  \nonumber
  \times
  \prod_{\sigma=\pm}
  \frac{\Gamma\!\left(
    \frac{1-\nu_B}{2}
    -\frac{i(\omega+\sigma m)}{2r_+}
  \right)}
  {\Gamma\!\left(
    \frac{1+\nu_B}{2}
    -\frac{i(\omega+\sigma m)}{2r_+}
  \right)}.
  \label{eq:exact-btz-response}
\end{align}

For positive timelike frequency $\omega>|m|$, with
$\omega\pm m\gg r_+$, Stirling expansion gives
\begin{equation}
  G_{R,-}^B(\omega,m)
  =\frac{2\nu_B\Gamma(1+\nu_B)}{\Gamma(1-\nu_B)}
   \left(\frac{2}{\sqrt{\omega^2-m^2}}\right)^{2\nu_B}
   e^{i\pi\nu_B}
   \left[1+\varepsilon_B(\omega,m)\right],
  \label{eq:btz-high-frequency}
\end{equation}
where
\begin{equation*}
  \varepsilon_B(\omega,m)
  =O\!\left(
    \frac{r_+^2}{(\omega-m)^2}
    +\frac{r_+^2}{(\omega+m)^2}
  \right).
\end{equation*}

 In order to write down the DT equation coupling the bath to the system, we need to convert the above cylinder basis into the AdS$_2$ Gegenbauer (defect) basis as in \cite{karch2026universalitydissipationholographicinterfaces}. Set $\nu_S=\nu_B=\nu$, and recall
$\lambda_k=\omega-k$ and $q^2=\lambda_k(\lambda_k-1)$.
We extend the response basis $v_{\nu_S,k}$ from \eqref{eq:source-response-profiles} by zero to
the bath full circle and define its Fourier coefficients by
\begin{equation}
  V_{m,k}=\int_{\mathcal H}\dd\varphi\,
  \frac{e^{-im\varphi}}{\sqrt{2\pi}}v_{\nu_S,k}(\varphi).
  \label{eq:fourier-profile}
\end{equation}
Using the normalization \eqref{eq:cross-normalization}, define
the effective bath response in the $k$th descendant channel
\begin{equation}
  G_{R,k}^{B,\mathrm{eff}}(\omega)
  =(2\lambda_k-1)\sum_{m\in\mathbb Z}
  V_{-m,k}G_{R,-}^B(\omega,m)V_{m,k}.
  \label{eq:effective-bath}
\end{equation}

We next evaluate it at large frequency limit and fixed $k$. In \eqref{eq:btz-high-frequency} we have the large $\omega$ fixed $m$ expression, and now we transform $m$ to $k$. The AdS$_2$ function is
$\chi_k\propto(\cos\varphi)^{\lambda_k}C_k^{(\lambda_k)}(\sin\varphi)$, and in the limit of large $\lambda_k$, set $z=\sqrt{\lambda_k}\,\varphi$, we have
\begin{equation*}
  \left(\cos\frac{z}{\sqrt{\lambda_k}}\right)^{\lambda_k}
  \longrightarrow e^{-z^2/2},
  \qquad
  \lambda_k^{-k/2}
  C_k^{(\lambda_k)}\!\left(
    \sin\frac{z}{\sqrt{\lambda_k}}
  \right)
  \longrightarrow\frac{H_k(z)}{k!}
\end{equation*}
Here $H_k$ is the Hermite polynomial. The main contribution then come from
$|\varphi|=O(\lambda_k^{-1/2})$, with its Fourier transformed range
\begin{equation}
  |m|=O(\lambda_k^{1/2})\ll\omega.
  \label{eq:localization}
\end{equation}
Recall that Parseval's identity and \eqref{eq:cross-normalization} then give
\begin{equation}
  (2\lambda_k-1)\sum_mV_{-m,k}V_{m,k}
  =1+O(\lambda_k^{-1}).
  \label{eq:packet-normalization}
\end{equation}

We still need to control the large $m$ region when we sum over the basis, but the upshot is that this region is negligible
\footnote{We thank GPT-Astra to point this out. To control the complementary Fourier tail, let
$Q=\operatorname{Re}\lambda_k\to\infty$, with
$\operatorname{Im}\lambda_k$ bounded.  For each fixed nonnegative integer
$J$, differentiation of the Gegenbauer profile and its Gaussian
localization give
\begin{equation}
  \left\|\partial_\varphi^J v_{\nu_S,k}\right\|_{L^2(\mathcal C)}^2
  \leq C_J Q^{J-1}.
  \label{eq:packet-sobolev}
\end{equation}
The norm contains complex conjugation and is used only to bound the
analytic bilinear.  Normalization contributes $Q^{-1}$ to the squared
norm, while each derivative contributes at most $Q^{1/2}$ to the amplitude.
Outside a fixed central angular interval,
$(\cos\varphi)^{\lambda_k-\nu_S-1}$ is exponentially suppressed;
its endpoint vanishing removes boundary terms in the first $J$
derivatives of the zero extension once $Q$ is sufficiently large.
These estimates are uniform on bounded complex-frequency neighborhoods
of the high levels.

For $0<\delta<1/2$, put $M_Q=Q^{1/2+\delta}$.  Fourier Parseval and
\eqref{eq:packet-sobolev} imply
\begin{equation}
  \sum_{|m|>M_Q}|V_{m,k}|^2
  \leq M_Q^{-2J}
     \left\|\partial_\varphi^Jv_{\nu_S,k}\right\|_{L^2}^2
  =O(Q^{-1-2J\delta}).
  \label{eq:packet-fourier-tail}
\end{equation}
We require the continued neighborhoods to remain a fixed positive
distance from the bath poles
\begin{equation}
  \omega=\pm m-ir_+(2p+1-\nu_B),
  \qquad m\in\mathbb Z,\quad p\in\mathbb N_0.
  \label{eq:bath-pole-separation}
\end{equation}
Real closed frequencies satisfy this condition because $0<\nu_B<1$.
On a bounded imaginary-frequency strip away from these poles, the exact
gamma ratios are bounded by constants times
$(1+|\operatorname{Re}\omega\pm m|)^{-\nu_B}$.
At least one light-cone frequency has magnitude of order $Q$ or larger,
so $\sup_m|G_{R,-}^B(\omega,m)|=O(Q^{-\nu_B})$.
Cauchy--Schwarz and \eqref{eq:packet-fourier-tail} therefore bound the
tail contribution to \eqref{eq:effective-bath} by
$O(Q^{-\nu_B-2J\delta})=o(Q^{-2\nu_B})$ when $2J\delta>\nu_B$.}. From \eqref{eq:btz-high-frequency}  and \eqref{eq:packet-normalization}, the descendant channel bath response is 
\begin{equation}
  G_{R,k}^{B,\mathrm{eff}}(\omega)
  =\frac{2\nu_B\Gamma(1+\nu_B)}{\Gamma(1-\nu_B)}
   \left(\frac2q\right)^{2\nu_B}
   e^{i\pi\nu_B}[1+o(1)].
  \label{eq:projected-bath}
\end{equation}

\subsection{The full double-trace equations}
\label{sec:double-trace}

We now construct the full mode matrix for the bottom-up geometry with
warp factor $A(u)$ in the optical coordinate $u$.  The construction follows \cite[sec.~3.3]{karch2026universalitydissipationholographicinterfaces}.
 In this subsection, we do not take the large-radial-level limit yet. In section~\ref{sec:poles} we will derive the solutions of the quasinormal modes in this limit, and we leave solving the full DT equations at finite frequency to future work.

When we couple the system asymptotic AdS to the BTZ bath, eigenfunctions with different AdS$_2$ channels $j$ will mix with each other with different strengths\footnote{Here we use $j$ for the AdS$_2$ descendant level and leave $k$ for the fixed channel number.}. Let us consider the most general system source/response with independent coefficients $C_j^S$ for the solution of the K-G equation
\begin{equation}
  \Phi^S=e^{-i\omega t}\sum_j C_j^S f_j(u;\omega)
                       \chi_j(\varphi;\omega).
  \label{eq:system-mode-superposition}
\end{equation}
The arbitrary normalization of each $f_j$ is absorbed into $C_j^S$.
From the individual coefficients in \eqref{eq:closed-massive-boundary-profiles}, the total source and
response in terms of the individual
descendant coefficients $a_j^S$ are given by
\begin{align}
  a^S(\omega,\varphi)
  &=\sum_j C_j^S a_j^S(\omega,\varphi)
   =\sum_j C_j^S\alpha_j^S s_{\nu,j}(\varphi),\nonumber\\
  b_R^S(\omega,\varphi)
  &=\sum_j C_j^S b_{j,R}^S(\omega,\varphi)\nonumber\\
  &=\sum_j C_j^S\left[
       \beta_j^S v_{\nu,j}(\varphi)
       +\frac{\alpha_j^S}{2\nu}
          \mathcal L_\nu(\omega)s_{\nu,j}(\varphi)\right].
  \label{eq:total-system-boundary-data}
\end{align}
Recall that $s_{\nu,j}=(\cos\varphi)^{\nu-1}\chi_j$ and
$v_{\nu,j}=(\cos\varphi)^{-\nu-1}\chi_j$, and we keep the 
$\mathcal L_\nu$ term to be consistent with \cite{karch2026universalitydissipationholographicinterfaces}.  

For ingoing bath mode amplitudes $C_m^B$, the corresponding sums are\footnote{As in the separated formulation of
\cite{karch2026universalitydissipationholographicinterfaces}, these
sums require a specified regulator or meromorphic
continuation.  For example, the response profile behaves as
$v_{\nu,j}\sim(\cos\varphi)^{\lambda_j-\nu-1}$. Its ordinary angular
integral converges for $\operatorname{Re}\lambda_j>\nu$, and it is in
$L^2(\mathcal H)$ for $\operatorname{Re}\lambda_j>\nu+1/2$.
These conditions fail for sufficiently large $j$ at fixed $\omega$.
The matrix of DT equations below will therefore be regulated separated representation of
the full boundary conditions.}
\begin{align}
  \Phi^B&=e^{-i\omega t}\sum_{m\in\mathbb Z}
          C_m^B\Phi_m^B(r;\omega)e_m(\varphi),\nonumber\\
  a^B(\omega,\varphi)&=\sum_m C_m^B a_m^B e_m(\varphi),
  \qquad
  b_R^B(\omega,\varphi)=\sum_m C_m^B b_{m,R}^B e_m(\varphi).
  \label{eq:total-bath-boundary-data}
\end{align}
In the following homogeneous DT equations we aim to solve the amplitudes $C_j^S$ and $C_m^B$ together with the frequency $\omega$.

Now we spell out the DT equations between system and bath. Define
the projection map $P_{\mathcal H}$ restricts bath functions from the full circle $\mathcal C$ to the half circle
$\mathcal H$, and $P_{\mathcal H}^*$ extends a system function by zero
to $\mathcal C$.  For a uniform real coupling $h_R$ the interaction is
\begin{equation}
  S_{\mathrm{int}}
  =-h_R\int_{\mathbb R_t}\dd t\int_{\mathcal H}\dd\varphi\,
    X_B\mathcal O_S.
  \label{eq:interaction-full}
\end{equation}
The source/response variations give the
reciprocal double-trace mixed conditions
\cite{Witten:2001ua,Papadimitriou:2007MultiTrace}
\begin{subequations}
\label{eq:mixed-bc}
\begin{align}
  a^S&=-2\nu h_R P_{\mathcal H}a^B,\\
  b_R^B&=2\nu h_R P_{\mathcal H}^*b_R^S.
\end{align}
\end{subequations}
Substituting the mode sums gives 
\begin{align}
  \sum_j C_j^S\alpha_j^S s_{\nu,j}
    +2\nu h_R\sum_m C_m^B a_m^B P_{\mathcal H}e_m&=0,\nonumber\\
  \sum_m C_m^B b_{m,R}^B e_m
    -2\nu h_R P_{\mathcal H}^*
        \sum_j C_j^S b_{j,R}^S&=0.
  \label{eq:full-functional-mode-equations}
\end{align}

Let us also choose a fixed complete orthonormal basis $u_\ell$ on $\mathcal H$ from
\cite[eq.~(3.35)]{karch2026universalitydissipationholographicinterfaces}.
Define the system source and response matrices and the restriction matrix by
\begin{align}
  \mathsf A_{\ell j}
    &=\int_{\mathcal H}\dd\varphi\,u_\ell^* a_j^S
     =\alpha_j^S\int_{\mathcal H}\dd\varphi\,
          u_\ell^*s_{\nu,j},\nonumber\\
  \mathsf V_{\ell j}
    &=\int_{\mathcal H}\dd\varphi\,u_\ell^* b_{j,R}^S\nonumber\\
    &=\beta_j^S\int_{\mathcal H}\dd\varphi\,u_\ell^*v_{\nu,j}
      +\frac{\alpha_j^S}{2\nu}
       \int_{\mathcal H}\dd\varphi\,
           u_\ell^*\mathcal L_\nu s_{\nu,j},\nonumber\\
  K_{\ell m}
    &=\int_{\mathcal H}\dd\varphi\,u_\ell^*e_m.
  \label{eq:system-mode-matrices}
\end{align}
  The adjoint $K^\dagger$ will represent zero extension
in these bases.  The bath coefficient matrices are
\begin{equation}
  \mathsf D_a=\operatorname{diag}_m(a_m^B),
  \qquad \mathsf D_b=\operatorname{diag}_m(b_{m,R}^B).
  \label{eq:bath-mode-matrices}
\end{equation}

Testing \eqref{eq:full-functional-mode-equations} with the basis $u_\ell^*$ and
$e_m^*$ gives the full coupled mode equation solving for the quasinormal modes
\begin{equation}
  \begin{pmatrix}
    \mathsf A(\omega)&2\nu h_R K\mathsf D_a(\omega)\\
    -2\nu h_R K^\dagger\mathsf V(\omega)&\mathsf D_b(\omega)
  \end{pmatrix}
  \begin{pmatrix}\boldsymbol C^S\\\boldsymbol C^B\end{pmatrix}=0.
  \label{eq:full-coupled-mode-matrix}
\end{equation}

Where $\mathsf D_b$ is invertible, we can further eliminate the bath amplitudes to give
\begin{equation}
\left[
    \mathsf A+4\nu^2h_R^2
       K\mathsf D_a\mathsf D_b^{-1}K^\dagger\mathsf V
  \right]\boldsymbol C^S=0.
  \label{eq:full-system-mode-matrix}
\end{equation}

In individual mode indices this is, for every $\ell$,
\begin{equation}
  \sum_j\left[
    \mathsf A_{\ell j}
    +4\nu^2h_R^2\sum_{m,\ell'}
      K_{\ell m}\frac{a_m^B}{b_{m,R}^B}
      K_{\ell' m}^*\mathsf V_{\ell' j}
  \right]C_j^S=0.
  \label{eq:full-system-mode-components}
\end{equation}
Together with \eqref{eq:system-mode-matrices}, this spells out the pole
equation in terms of $\alpha_j^S,\beta_j^S$.

Let us connect to the notation of the system and bath response notations. We have
$\mathsf G_B=\operatorname{diag}_m G_{R,-}^B(\omega,m)
=-2\nu\mathsf D_a\mathsf D_b^{-1}$.
Then \eqref{eq:full-system-mode-matrix} is
$(\mathsf A-2\nu h_R^2K\mathsf G_BK^\dagger\mathsf V)
\boldsymbol C^S=0$.
If $\mathsf A$ is invertible, the system response is
$\mathsf G_S=2\nu\mathsf V\mathsf A^{-1}$ and the coefficients of
the total source in the $u_\ell$ basis are
$\boldsymbol a^S=\mathsf A\boldsymbol C^S$.  Thus the equivalent
DT equation is \cite{karch2026universalitydissipationholographicinterfaces,vargas2026holographicopenquantum}
\begin{equation}
  \left[\mathbf1_{\mathcal H}
    -h_R^2P_{\mathcal H}G_{R,-}^BP_{\mathcal H}^*G_R^S\right]a^S=0.
  \label{eq:exact-operator-equation}
\end{equation}

We will solve this in Section~\ref{sec:poles} where we turn to its simpler large-$n$, fixed-$k$
projection.

\section{The system response and BOPE coefficients}
\label{sec:closed-bope}

The leading order system Green function in \eqref{eq:system-response-leading} becomes subtle near each normal mode pole of the holographic ICFT. It does not yet give its the residue $\mathcal{R}_{n,k}^S$. The residue is of physical interest since it connects
the spectra and responses of those closed normal modes to defect correlation functions, or specifically the BOPE coefficients.
In this section we show that the optical length $\rho=L/\pi$, which governs the defect spectrum at large dimension, also determines the behavior of BOPE coefficients of the conformal interface in this limit.

\subsection{BOPE coefficients in the WKB limit}
\label{sec:exact-pole-weight}

At a positive normal mode pole $\Omega_{n,k}=d_n+k$, we have
\begin{equation}
  G_{R,k}^{S,\mathrm{proj},(\nu)}(\omega)
  =-\frac{\mathcal R_{n,k}^S}{\omega-\Omega_{n,k}}
    +G_{n,k}^{S,\mathrm{reg}}(\omega),
  \qquad
  \mathcal R_{n,k}^S\equiv
  -\Res_{\omega=\Omega_{n,k}}G_{R,k}^{S,\mathrm{proj},(\nu)}(\omega)>0 .
  \label{eq:projected-pole-definition}
\end{equation}
Here $G_{n,k}^{S,\mathrm{reg}}$ is analytic at the selected pole. In the following we express this residue in terms of the exact scalar solutions from \eqref{eq:exact-solution}.

Consider the exact right-normalized solution \(\Psi_R(u;q)\) in \eqref{eq:exact-solution} and its decomposition \eqref{eq:leftrightexact} in terms of the left endpoint basis. At a normal mode frequency \(\Omega_{n,k}=d_n+k\), we set
$E_k(\Omega_{n,k})=q_n^2=d_n(d_n-1)$. Since $\alpha_k^S(\Omega_{n,k})=0$, the two normalized solutions are therefore proportional:
\begin{equation}
    \Psi_R(u;q_n)
=\beta_k^S(\Omega_{n,k})\Psi_L(u;q_n).
\end{equation}

The residue is determined by the derivative of the source coefficient. Let us compare the fixed eigenfunction \(\Psi_L(u;q_n)\) with \(\Psi_R(u;q)\) at a nearby frequency. Their  radial K-G equations with the same potential \(V_\Delta(u)\) imply
\begin{equation}
    \partial_u\mathcal W[\Psi_L(u;q_n),\Psi_R(u;q)]
=
\bigl[q_n^2-E_k(\omega)\bigr]
\Psi_L(u;q_n)\Psi_R(u;q),
\end{equation}

 The right boundary term vanishes since both functions obey the source-free boundary condition there. At the left side, the source coefficient is related to the Wronskian by definition:
\[
\lim_{u\to0}
\mathcal W[\Psi_L(u;q_n),\Psi_R(u;q)]
=-2\nu\alpha_k^S(\omega).
\]

Integration over the full interval consequently gives
\[
2\nu\alpha_k^S(\omega)
=
\bigl[q_n^2-E_k(\omega)\bigr]
\int_0^L\Psi_L(u;q_n)\Psi_R(u;q)\,du.
\]

Define the norm of the real, left-normalized eigenfunction by
\[
\mathcal I_n=\int_0^L\Psi_L(u;q_n)^2\,du.
\]

Differentiating with respect to \(\omega\) at fixed \(k\), and evaluating at \(\Omega_{n,k}\), yields
\[
2\nu\,\partial_\omega\alpha_k^S(\Omega_{n,k})
=
-(2d_n-1)\beta_k^S(\Omega_{n,k})\mathcal I_n,
\]
where we used $\left.\partial_\omega E_k(\omega)\right|_{\omega=\Omega_{n,k}}
=2d_n-1$. Then we have
\begin{equation}
\operatorname*{Res}_{\omega=\Omega_{n,k}}
G_{R,k}^{S,\mathrm{proj},(\nu)}(\omega)
=
\frac{2\nu\beta_k^S(\Omega_{n,k})}
{\partial_\omega\alpha_k^S(\Omega_{n,k})}
=
-\frac{4\nu^2}{(2d_n-1)\mathcal I_n}.
\end{equation}

Notice that this indicates that the residue is actually independant of $k$:

\begin{equation}
  \mathcal R_{n,k}^S=\mathcal R_n^S
  \equiv-\operatorname*{Res}_{\omega=\Omega_{n,k}}
       G_{R,k}^{S,\mathrm{proj},(\nu)}(\omega)
  =\frac{4\nu^2}{(2d_n-1)\mathcal I_n}.
  \label{eq:real-axis-norming-residue}
\end{equation}

Going into the large $n$ fixed $k$ limit, we consider the interval $[\epsilon,L-\epsilon]$ where we did the Bessel matching
in \eqref{eq:closed-interior-solutions}:
\begin{equation}
  \Psi_L(u;q_n)=C_\nu(q_n)
  \left[\cos\left(q_nu-\frac{\pi\nu}{2}-\frac\pi4\right)
  +O(q_n^{-1})\right],
  \qquad
  C_\nu(q)=2^\nu\Gamma(\nu+1)\sqrt{\frac2\pi}\,q^{-\nu-1/2}.
  \label{eq:real-axis-interior-amplitude}
\end{equation}
The interior contribution to $\mathcal I_n/C_\nu(q_n)^2$ therefore tends
to $(L-2\epsilon)/2$. We further control the two endpoint regions where the integral is bounded by $O(\epsilon C_\nu(q_n)^2)$ \cite{KostenkoSakhnovichTeschl:2010Inverse,KostenkoTeschl:2011SingularWeyl}. As usual we first take $n\to\infty$ and then $\epsilon\to0$, this gives the leading order behavior of the integral
\begin{equation}
  \mathcal I_n=\frac L2 C_\nu(q_n)^2
  =\frac{2^{2\nu}\Gamma(\nu+1)^2L}
         {\pi q_n^{2\nu+1}}[1+o(1)].
  \label{eq:real-axis-endpoint-norm}
\end{equation}

Substituting into \eqref{eq:real-axis-norming-residue} gives the residue
\begin{equation}
\mathcal R_n^S
  =\frac{2}{\Gamma(\nu)^2\rho}
     \left(\frac{q_n}{2}\right)^{2\nu}[1+o(1)],
  \label{eq:massive-bope-WKB}
\end{equation}

\subsection{The BOPE coefficients}
\label{sec:bope-normalization}

In interface/defect CFTs, an important set of data of a probe operator is its BOPE correlation functions, specifically the BOPE coefficients. Let us translate the above residue of response functions into the BOPE data.

Denote
$b_n\equiv b_{\mathcal O_S\widehat{\mathcal O}_n}$ be the BOPE
coefficient of the system probe scalar, with conventions
\begin{equation*}
  \mathcal O_S(t,x_\perp)
  =\sum_n b_n(2x_\perp)^{d_n-\Delta}
        \widehat{\mathcal O}_n(t)+\text{descendants},
\end{equation*}
where the defect primaries $\widehat{\mathcal O}_n$ has the normalization $\langle\widehat{\mathcal O}_n(t)\widehat{\mathcal O}_n(0)\rangle
  =|t|^{-2d_n}$ and $x_\perp>0$ measures distance from the interface \cite{McAvityOsborn:1995BoundaryCFT,LiendoRastelliVanRees:2013BCFT,
BilloEtAl:2016Defects}. Recall
$\Delta\equiv\Delta_S=1+\nu$. For 2D ICFT, the relation between the residue of response functions at each pole and the BOPE coefficients were studied in \cite{Chiodaroli:2016jod,karch2026universalitydissipationholographicinterfaces} for BPS-Janus. Here let us generalize to our example.

Write one contribution
to the full cylinder kernel as
\begin{equation}
  G_R^S(\omega;\varphi,\varphi')\supset
  \frac{\mathcal W_{n,k}\,
    \widehat\chi_{n,k}(\varphi)\widehat\chi_{n,k}(\varphi')}
       {(d_n+k)^2-(\omega+i0)^2},
\end{equation}
for the normalized functions $\widehat\chi_{n,k}(\varphi)
  =(\cos\varphi)^{d_n-\Delta}C_k^{(d_n)}(\sin\varphi)$. The $SL(2,\mathbb{R})$ conformal descendant expansion of $\widehat{\mathcal O}_n$ relates $\mathcal{W}$ to the BOPE coefficients by\footnote{For a unit-normalized defect primary of dimension $d_n$,
the descendants have norms $\|P^k|n\rangle\|^2=k!(2d_n)_k$.
Expanding the bulk--defect 2-pt function on the cylinder with
the Gegenbauer generating function gives the normalized matrix-element
coefficient $2^{d_n-\Delta}b_n\sqrt{k!/(2d_n)_k}$ multiplying
$\widehat\chi_{n,k}$. Squaring this coefficient and Fourier transforming
$e^{-(d_n+k)|\tau|}$ gives
$\mathcal W_{n,k}/[2(d_n+k)]$, yielding the stated relation.}
\begin{equation}
  \frac{\mathcal W_{n,k}}{2(d_n+k)}
  =2^{2(d_n-\Delta)}
       \frac{k!\Gamma(2d_n)}{\Gamma(2d_n+k)}b_n^2.
  \label{eq:massive-bope-descendant}
\end{equation}
Comparing $\widehat\chi_{n,k}$ with the on-shell function
$\chi_{n,k}=\chi_k(\Omega_{n,k})$ in \eqref{eq:ads2eigen}
gives
\begin{equation}
  \begin{aligned}
  \widehat\chi_{n,k}(\varphi)
  &=\mathcal N_{n,k}(\cos\varphi)^{-\Delta}\chi_{n,k}(\varphi),\\
  \mathcal N_{n,k}^2
  &=\pi2^{2-2d_n}
     \frac{\Gamma(2d_n+k)}{k!\Gamma(d_n)^2}.
  \end{aligned}
  \label{eq:massive-bope-profile}
\end{equation}
The source profile is
$s_{\nu,k}=(\cos\varphi)^{\Delta-2}\chi_{n,k}$.  Using
\eqref{eq:channel-normalization} on each side of the kernel, the
projected pole residue becomes
\begin{equation}
  \mathcal R_n^S
  =\frac{\mathcal W_{n,k}}{2(d_n+k)}(2d_n-1)
       \left[\int_{\mathcal H}s_{\nu,k}\widehat\chi_{n,k}
                    \,\dd\varphi\right]^2
  =\frac{\mathcal W_{n,k}}{2(d_n+k)}
       \frac{\mathcal N_{n,k}^2}{2d_n-1}.
  \label{eq:massive-bope-projection-conversion}
\end{equation}

Picking out the factor we have
\begin{equation}
  \mathcal R_n^S=\mathcal K_\Delta(d_n)b_n^2,
  \qquad
  \mathcal K_\Delta(d)=
  \frac{\pi2^{2-2\Delta}\Gamma(2d)}
       {(2d-1)\Gamma(d)^2}>0\quad(d>1/2).
  \label{eq:massive-bope-dictionary}
\end{equation}.  

This also reproduces the coefficient
$(2d_n-1)\mathcal R_n^S$ of the full-kernel pole in
\eqref{eq:full-kernel-pole-contribution}. 

In the large dimension limit of \eqref{eq:massive-bope-WKB}, the leading order behavior of BOPE coefficients $b_I(d_n)\equiv b_n$ in this limit in terms of the defect primary dimension is
\begin{equation}
  b_I^2(d)
  =\frac{2(2d-1)\Gamma(d)^2[d(d-1)]^{\Delta-1}}
  {\pi\Gamma(\Delta-1)^2\Gamma(2d)\rho}\sim\frac{2^{3-2d}}
  {\sqrt\pi\,\Gamma(\Delta-1)^2\rho}
       d^{2\Delta-3/2}.
  \label{eq:massive-bope-envelope-asymptotic}
\end{equation}

 We check the above normalizations and BOPE coefficients in the pure-$\AdS_3$ reference limit in Appendix~\ref{app:pure-ads}. In addition, in Appendix~\ref{app:tauberian} we use the Tauberian theorem for 1D conformal blocks in \cite{QiaoRychkov2017} to check the above factor of $\mathcal{K}_\Delta$. Comparing the two cases with conformal interface $I$ and without it at the same defect dimension $d$ gives
\begin{equation}
  \lim_{d\to\infty}\frac{b_{I}^2(d)}
       {b_{\AdS}^2(d)}=\frac1\rho.
  \label{eq:massive-bope-ratio}
\end{equation}

This is another main result of this paper: In additional to the defect primary spectra, the BOPE coefficients of the operator in the same large radial dimension limit is also governed by the optical length $\rho$, which also has the nice geometric interpretation in the bottom-up 3D holographic model.

\section{Dissipation rate and $c_{UV}$}
\label{sec:poles}

We now return to calculating the system+bath marginal DT coupling in Section \ref{sec:uniform-btz}. In this section, we couple the fixed descendant channel to the BTZ bath at large defect dimension
and calculate the leading behavior of the dissipation introduced to the system by the bath. We then introduce the central-charge like quantity $c_{UV}$ in \cite{karch2026universalitydissipationholographicinterfaces} determined by the ratio of the dissipation rate in this limit, and prove that yet again it is related to the same geometric quantity $\rho$.

\subsection{The dissipation rate in the WKB limit}
\label{sec:projected-opening}

Let us start from the DT equation \eqref{eq:exact-operator-equation} that we repeat here:
\[
  \left[\mathbf1_{\mathcal H}
    -h_R^2P_{\mathcal H}G_{R,-}^B
       P_{\mathcal H}^{*}G_R^S\right]a^S=0.
\]
Let us project it to the fixed-$k$th channel, by substituting the source with
$s_{\nu,k}$ and doing the inner product with the response basis $v_{\nu,k}$.
Using the normalization relation \eqref{eq:cross-normalization} and the projected response \eqref{eq:full-response-on-channel}, the DT equation in the large-$n$ fixed-$k$ WKB limit is then\footnote{Here we drop the local contact terms. Explicitly it is $\mathcal E_k^{\mathrm{loc}}
      =-h_R^2(2\lambda_k-1)
        \langle v_{\nu,k},P_{\mathcal H}G_{R,-}^B
                  P_{\mathcal H}^{*}\ell_k^{\mathrm{loc}}\rangle.$ where $\ell_k^{\mathrm{loc}}
      =\mathcal L_\nu s_{\nu,k}
          -G_{\mathrm{loc}}^{(\nu)}v_{\nu,k}$. Also note that this is for nonzero coupling. At $h_R=0$, the appropriate equation for the normal poles is instead
\begin{equation*}
  F_k(\omega)
  =\alpha_k^S\left[
       1+\mathcal E_k^{\mathrm{loc}}
       -h_R^2G_{\mathrm{loc}}^{(\nu)}G_{R,k}^{B,\mathrm{eff}}\right]
    -2\nu h_R^2\beta_k^S G_{R,k}^{B,\mathrm{eff}}=0,
  \label{eq:projected-mode-numerator}
\end{equation*}
and at $h_R=0$ it reduces to $\alpha_k^S=0$.}
\begin{align}
  D_k(\omega)
  &\equiv(2\lambda_k-1)
    \left\langle v_{\nu,k},
    \left[\mathbf1_{\mathcal H}
      -h_R^2P_{\mathcal H}G_{R,-}^B P_{\mathcal H}^{*}G_R^S\right]
      s_{\nu,k}\right\rangle \nonumber\\
  &=1-h_R^2G_{R,k}^{S,\mathrm{proj},(\nu)}(\omega)
             G_{R,k}^{B,\mathrm{eff}}(\omega)=0,
  \label{eq:projected-scalar-equation}
\end{align}

where we also used
$G_{R,k}^{B,\mathrm{eff}}=(2\lambda_k-1)\langle v_{\nu,k},P_{\mathcal H}G_{R,-}^B
P_{\mathcal H}^{*}v_{\nu,k}\rangle$. 

We can finally use the expression of the projected responses in leading order obtained in
\eqref{eq:system-response-leading} and \eqref{eq:projected-bath} to derive the WKB poles of the coupled system+bath. Let us repeat their expressions here and analytically continue them into the complex domain where the poles will sit due to dissipation:
\begin{equation}
  G_{R,k}^{S,\mathrm{proj},(\nu)}(\omega)
  =-\frac{2\pi}{\Gamma(\nu)^2}
    \left(\frac q2\right)^{2\nu}
    \left[\cot\Theta_{\nu,k}(\omega)
            +\cot(\pi\nu)\right],
  \label{eq:continued-system-response}
\end{equation}
where we recall that the normal poles are at $\Theta_{\nu,k}(\Omega_{n,k})=n\pi$. The bath projected response is
\begin{equation}
  G_{R,k}^{B,\mathrm{eff}}(\omega)
  =\frac{2\nu\Gamma(1+\nu)}{\Gamma(1-\nu)}     \left(\frac2q\right)^{2\nu}e^{i\pi\nu}.
  \label{eq:continued-opening-estimates}
\end{equation}

Set $\mathfrak g=4\nu^2h_R^2$, the DT equation \eqref{eq:projected-scalar-equation} at a fixed nonzero coupling at leading order is simplified to
\begin{equation}
  1+\mathfrak g\sin(\pi\nu)e^{i\pi\nu}
    \left\{\cot\Theta_{\nu,k}(\omega)             +\cot(\pi\nu)\right\}=0.
  \label{eq:matched-scalar-secular}
\end{equation}

Using $\sin(\pi\nu)e^{i\pi\nu}=[\cot(\pi\nu)-i]^{-1}$,
the leading equation becomes
\begin{equation}
  e^{2i[\Theta_{\nu,k}(\omega)-n\pi]}
  =\frac{1+\mathfrak g}{1+\mathfrak g e^{2\pi i\nu}}.
  \label{eq:matched-finite-secular}
\end{equation}
Recall from the definition of $\Theta$ and the large dimension defect spectrum, we had
$\Theta_{\nu,k}(\omega)-n\pi
=L(\omega-\Omega_{n,k})+o(1)$ for large $n$.
Then the shifted poles from the normal modes satisfy
\begin{equation}
  \omega_{n,k}^{\mathrm{open,proj}}-\Omega_{n,k}
  =\frac{1}{2\pi i\rho}
    \log\!\left(
       \frac{1+\mathfrak g}{1+\mathfrak g e^{2\pi i\nu}}
    \right)+o(1).
  \label{eq:matched-finite-shift}
\end{equation}
The $n\to\infty$ limit of the dissipation rate $\gamma\equiv -\Im \omega$ is then extracted from its imaginary part\footnote{In this paper we do not write down the subleading corrections at large $n$ or large $q$ carefully. For that purpose one need to do a more detailed analysis of the WKB approximation for the K-G solutions. We thank GPT-Astra for performing this calculation and report the results here. For fixed $k$, we use the reference potential \eqref{eq:bounded-endpoint-remainder} and the near-boundary comparison potential \eqref{eq:closed-Bessel-endpoint}. The result is
\begin{equation}
  \gamma_{n,k}^{\mathrm{proj}}
  =\gamma_\infty+\frac{1}{q_n^2}
  \left[
    \frac{\zeta_\nu\,\mathfrak g(1-\mathfrak g)\sin^2(\pi\nu)}
    {L(1+\mathfrak g)
     [1+2\mathfrak g\cos(2\pi\nu)+\mathfrak g^2]}
    -\eta_\nu\gamma_\infty
  \right]+o(q_n^{-2}),
\end{equation}
where $\gamma_\infty$ is the above UV limit,
$\eta_\nu=\tfrac18+\tfrac12\overline W_\nu$, and
$\zeta_\nu=-\nu v_{\nu,\mathrm{end}}
+\tfrac13\nu(1-\nu^2)r_+^2$.
Here 
\begin{align*}
  W_\nu(u)
  &\equiv V_\Delta(u)
    -\left(\nu^2-\frac14\right)
      \left(\frac{\pi}{L}\right)^2
      \csc^2\!\left(\frac{\pi u}{L}\right),
  &
  \overline W_\nu
  &\equiv \frac1L\int_0^L \dd u\,W_\nu(u),
\end{align*}

 and
\begin{equation*}
  v_{\nu,\mathrm{end}}
  \equiv \lim_{y\to0^+}
  \left[
    V_\Delta(y)-\frac{\nu^2-\tfrac14}{y^2}
  \right],
\end{equation*}
where $y$ is $u$ or $L-u$. The subleading correction is then of order
$n^{-2}$ and is independent of $k$.}:
\begin{equation}
  \gamma_{\infty}^{\mathrm{proj}}
  =\frac{1}{2\pi\rho}
    \log\!\left(
       \frac{1+\mathfrak g}
            {|1+\mathfrak g e^{2\pi i\nu}|}
    \right).
  \label{eq:matched-finite-width}
\end{equation}

One can check that it is positive for $\mathfrak g>0$ since
\[
  (1+\mathfrak g)^2-|1+\mathfrak g e^{2\pi i\nu}|^2
  =4\mathfrak g\sin^2(\pi\nu)>0.
\]

This is different from the dissipation rate found in \cite{karch2026universalitydissipationholographicinterfaces} for the massless scalar in BPS-Janus: We have a nonzero dissipation rate in the infinite defect dimension limit for the general massive scalar for bottom-up model, whereas for the massless scalar in BPS-Janus it was logarithmically approaching zero.

It is also interesting to point out\footnote{We thank Andreas Karch for bringing the author's attention to this point.} that the above expression in terms of the DT coupling is analogous to the optical depth theorem for the no-interface case DT coupling between pure AdS$_3$ and BTZ found in \cite{vargas2026holographicopenquantum}. The RHS of \eqref{eq:matched-finite-width} in the pure AdS$_3$ case is closely related to the energy transmission between the system and bath \cite{karch2025dissipationopenholography}. We leave its physical interpretation in presence of the conformal interface to future work \cite{future}.

\subsection{$c_{UV}$}
\label{sec:uv-diagnostic}
\label{sec:bath-universality}

As in \cite{karch2026universalitydissipationholographicinterfaces} we define a ratio between the dissipation rate in this projected UV limit with and without and interface at the same nonzero coupling $h_R$. The reference pure AdS$_3$ has
$L_{\AdS}=\pi$ and $\rho_{\AdS}=1$. We now define
\begin{equation}
  \frac{\cUV}{c}
  \equiv
    \frac{\gamma_{\mathrm{int},\infty}^{\mathrm{proj}}(h_R)}
         {\gamma_{\AdS,\infty}^{\mathrm{proj}}(h_R)}.
  \label{eq:cuv-definition}
\end{equation}
For the uniform marginal DT coupling and the massive probe scalars, we have:
\begin{equation}
\frac{\cUV}{c}=\frac1\rho=\frac{\pi}{L}.
  \label{eq:cuv-density}
\end{equation}

Yet again, the open interface quantity $c_{UV}$ is related to the same optical length in the projected UV limit as the BOPE coefficients:
\begin{equation}
\frac{\cUV}{c}=\lim_{d\to\infty}\frac{b_{I}^2(d)}
       {b_{\AdS}^2(d)}=\frac1\rho.
  \label{eq:cuv-bope-summary}
\end{equation}

We thus enhance the universality of $c_{UV}$ from \cite{karch2026universalitydissipationholographicinterfaces} for general probe scalars in generic bottom-up holographic ICFTs\footnote{We can also consider the massless case in \cite{karch2026universalitydissipationholographicinterfaces} for $\nu=1$. As before, we use the dynamical-source
prescription in
\cite{karch2026universalitydissipationholographicinterfaces} with fixed
subtraction scales $\mu_S,\mu_B$. 
The system response in \eqref{eq:closed-massive-integer-response} and
the renormalized bath response become
\begin{align}
  G_{R,k}^{S,\mathrm{proj},(1)}
  &=-\frac{q^2}{2}
    [\pi\cot\Theta_{1,k}(\omega)
                      +\ell_S(q)+o(1)],\nonumber\\
  G_{R,k}^{B,\mathrm{eff},(1)}
  &=\frac{8}{q^2[\ell_B(q)-i\pi+o(1)]},
  \label{eq:massless-renormalized-kernels}
\end{align}
where the local contact terms are crucial for subtracting the divergences and
\[
  \ell_S(q)=2\log\frac{q}{2\mu_S}+O(1),\qquad
  \ell_B(q)=2\log\frac{q}{2\mu_B}+O(1).
\]
  Under these conditions,
\begin{equation*}
  \pi \cot\Theta_{1,k}(\omega)
  =-\frac{\ell_B(q)+4h_R^2\ell_S(q)-i\pi+o(1)}{4h_R^2}.
  \label{eq:massless-log-secular}
\end{equation*}
Set $A_n=\ell_B(q_n)+4h_R^2\ell_S(q_n)$. Its leading behavior is
$A_n=2(1+4h_R^2)\log(q_n/\mu_B)+O(1)$ for fixed subtraction scales.
Linearizing about that real solution, using
$\partial_\omega\Theta_{1,k}=L+o(1)$ gives
\begin{align*}
  \gamma_{n,k}^{\mathrm{proj}}
  \sim\frac{\pi h_R^2}
   {\rho(1+4h_R^2)^2\log^2(q_n/\mu_B)}.
  \label{eq:massless-fixed-coupling-width}
\end{align*}
 $c_{UV}$ for the massless case is defined by the ratio of $\gamma_{n,k}$ at fixed large $q$ and fixed $k$, taking $q\to\infty$ at the end. We have again $c_{UV}/c=1/\rho$.}.

\section{A new holographic bound}
\label{sec:new-bound}

We now use the geometric expression for $\cUV$ in the bottom-up model to derive a new holographic bound
on energy transmission $c_{LR}$ in terms of $\ceff$ and $\cUV$. In \cite{Karch:2024udk} it was proved that $c_{LR}\leq c_\text{eff}$ for the most general bottom-up holographic interfaces, whose bulk warpfactor satisfies
the null energy condition (NEC). In this section, we prove a lower bound of $c_{LR}$ by $c_{UV}$ and $c_\text{eff}$ for the slightly constrained warpfactors described in Section \ref{sec:closed-modes}.

First let us recall the energy transmission $c_{LR}$ \cite{Quella_2007} and entanglement entropy $c_{\text{eff}}$ \cite{Sakai_2008,Karch:2023evr} for closed conformal interfaces in 2D CFT. $c_{LR}$ is fixed by the 2-pt function of stress tensors across the interface, while $c_\text{eff}$ is given by the universal logarithm coefficient in entanglement entropy across the interface.

In bottom-up holography they both have a nice interpretation in terms of bulk geometry. We keep the common unit asymptotic
$\AdS_3$ radius, and ambient central charges 
$c=3/(2G_N)$. Let $\Amin=\min_x A(x)$ and define $I_2$ along with our optical length $\rho$:
\begin{equation}
  \rho=\frac1\pi\int_{-\infty}^{\infty}\frac{\dd x}{A(x)},
  \qquad I_2=\int_{-\infty}^{\infty}\frac{\dd x}{A(x)^2}.
  \label{eq:bound-functionals}
\end{equation}

In this setup $c_{LR}$, $c_\text{eff}$ and the open interface quantity $c_{UV}$ at hand are given by \cite{BachasEtAl:2023ThickBranes,Karch_2021}
\begin{equation}
  \ceff=c\Amin,
  \qquad \cLR=\frac3{G_N I_2},
  \qquad \cUV=\frac c\rho.
  \label{eq:bound-observables}
\end{equation}

 In terms of the optical coordinate $u$ from \eqref{eq:closed-optical-data} and denote $B(u)=1/A(x(u))$,
the interval has length $L=\pi\rho$ and we can rewrite the definition
\begin{equation}
  I_2=\int_0^L B(u)\,\dd u.
  \label{eq:bound-I2-optical}
\end{equation}

We will use these expressions to come up with the new bounds below. 

\subsection{The NEC and the bound on \texorpdfstring{$\cUV$}{cUV}}
\label{sec:bound-NEC}

We first establish some basics regarding $c_{UV}$. For the closed interface quantities $c_{LR}$ and $c_\text{eff}$, they describe how the conformal interface suppresses energy and quasi-particle entanglement transmission, and it is physically natural that $0\le c_{LR},c_\text{eff}\le c$ (proved in \cite{Meineri_2020,Karch:2023evr}). For $c_{UV}$ it is far less obvious. Therefore, we first prove that $0\le c_{UV}\le c$ is also true, making it another central-charge like quantity measuring how the interface suppresses UV dissipation. 

For the bottom-up metric in \eqref{eq:closed-metric}, the null energy condition on the holographic bulk is $1+(A'(x))^2-A(X)A''(x)\geq0$. In $B(u)$ it is
\begin{equation}
  B''(u)+B(u)\geq0.
  \label{eq:bound-optical-NEC}
\end{equation}
Its boundary conditions are
\begin{equation}
  B(0)=B(L)=0,
  \qquad B'(0)=1,\qquad B'(L)=-1,
  \label{eq:bound-endpoint-data}
\end{equation}


To prove $L\geq\pi$, consider the integral on $[0,L]$:
\begin{equation}
  0\leq\int_0^L (B(u)+B''(u))\sin\!\left(\frac{\pi u}{L}\right)\dd u
  =\left(1-\frac{\pi^2}{L^2}\right)
    \int_0^L B(u)\sin\!\left(\frac{\pi u}{L}\right)\dd u.
  \label{eq:bound-optical-length}
\end{equation}
The last integral is positive, so $L\geq\pi$. The saturation condition is that NEC is saturated everywhere, giving pure AdS$_3$ with transparent interface. Hence we have
\begin{equation}
\qquad 0<c_{UV}\leq c.
  \label{eq:cuv-baseline}
\end{equation}

\subsection{The lower bound on transmission}
\label{sec:bound-cap}

Set $M\equiv B_\text{max}=1/A_\text{min}$. Consider the upper bound of $I_2$.
 A naive estimate is $I_2\leq ML$, but we can also use NEC to improve it\footnote{See Appendix \ref{app:geometric-proof} for the proof. We thank GPT-Astra for pointing this out.}:
\begin{equation}
  \int_0^L[M-B(u)]\,\dd u\geq M(\pi-2),
  \label{eq:bound-cap-deficit}
\end{equation}
or equivalently
\begin{equation}
  I_2\leq M(L-\pi+2).
  \label{eq:bound-I2-sharp}
\end{equation}

Substituting $M=c/\ceff$ and $L=\pi c/\cUV$ gives our new lower bound on $c_{LR}$:
\begin{equation}
  \cLR\geq
  \frac{2\ceff}{2+\pi\left(\frac c{\cUV}-1\right)}.
  \label{eq:bound-main}
\end{equation}
Together with its upper bound in \cite{Karch:2024udk}, the bounds for bottom-up holographic ICFT are
\begin{equation}
  0<\frac{2\ceff}{2+\pi(c/\cUV-1)}
  \leq\cLR\leq\ceff\leq c.
  \label{eq:final-bound}
\end{equation}

\subsection{Saturation of the bound}
\label{sec:bound-saturation}

When the lower bound saturates, equality in \eqref{eq:bound-cap-deficit} requires $B=M$ except for AdS$_3$ caps near the boundaries, since saturation of NEC is stronger. In addition we need $M=1$, or $\ceff=c$.  In the original slicing coordinate, the
complete equality family is, up to translation, (see Appendix~\ref{app:geometric-proof} for details)
\begin{equation}
  A(x)=
  \begin{cases}
    \cosh x, &x\leq0,\\
    1, &0\leq x\leq L-\pi,\\
    \cosh(x-L+\pi), &x\geq L-\pi,
  \end{cases}
  \qquad L\geq\pi.
  \label{eq:bound-saturating-warp-factor}
\end{equation}
It consists of two vacuum $\AdS_3$ caps joined by an
$\AdS_2\times\mathbb R$ throat of proper length $L-\pi$, with 
\begin{equation}
  \ceff=c,\qquad
  \frac{\cUV}{c}=\frac\pi L,\qquad
  \frac{\cLR}{c}=\frac{2}{L-\pi+2}.
\end{equation}

In Appendix~\ref{app:janus-examples} we also verify the commonly used holographic Janus interfaces against the new bound. A byproduct of our result is that the coincidence $c_{UV}=c_\text{eff}$ found in \cite{karch2026universalitydissipationholographicinterfaces} for BPS-Janus is not general and special for that example.

\section{Conclusion and outlook}
\label{sec:discussion}

In this paper we related three things together in the bottom-up holographic ICFT for a probe scalar in the large-defect-dimension limit fixing the descendant level: its defect primary spectrum, BOPE coefficients, and 
dissipation rate under uniform DT coupling to an external bath. In our description, this limit corresponds to the radial WKB regime. Under mild assumptions on the warpfactor $A(x)$, we proved

\begin{equation*}
  \frac{\cUV}{c}
  =\lim_{d\to\infty}\frac{b_I^2(d)}{b_{\AdS}^2(d)}
  =\frac1\rho
  =\frac{\pi}{\displaystyle\int_{-\infty}^{\infty}\dd x/A(x)}.
\end{equation*}

This extends
the massless BPS-Janus result of
\cite{karch2026universalitydissipationholographicinterfaces}. In addition, the new bound \eqref{eq:final-bound} on interface observables relates open system dissipation to entanglement and energy
transmission of the closed conformal interface.

There are several directions to explore for future work. An immediate followup would be to calculate the full DT quasinormal modes in \eqref{eq:full-coupled-mode-matrix} and consider the analogous of $c_\text{relax}$ in \cite{barad2025dissipationmeetsconformalinterface} for holographic interfaces. In \cite{karch2026universalitydissipationholographicinterfaces}, the projected UV dissipation rate goes to zero, while here for massive scalars it goes to a positive number. This suggests a nontrivial Liouvillian gap for the system, provided that the full set of quasinormal modes behave similarly at large $\omega$. It would also be interesting to study 
nonmarginal and nonuniform couplings
to further enhance the universality of these open-system quantities. More recently, it was also pointed out in \cite{barad2026universalsuppressiondissipationconformal} that boundary conditions of fields and global structure of conformal interfaces (in the sense of \cite{Bachas_2002}) can make these open-system quantities ambiguous. This motivates us to track individual quasinormal-mode branches and examine how their relaxation rates connect to the projected ultraviolet \(c_{\mathrm{UV}}\) defined here.

Another direction is to explore open holographic CFT coupling to external baths in general dimensions using field theory techniques, such as the gravitational Schwinger-Keldysh (grSK) formulism developed in \cite{Jana_2020,Loganayagam_2023,ishii2025lindbladdynamicsholography}. At the end of the day, open holographic systems are very different from the Markovian Lindbladian dynamics considered in \cite{barad2026universalsuppressiondissipationconformal}. A fruitful route is to first study the thermal observables/responses of a gravitational bath, then use the grSK formulism to study open fields coupled to them by integrating the bath out. In addition, one may consider extending the scope of \cite{Bonelli_2022,jia2024holographicthermalcorrelatorsquasinormal} to relate the bath correlators in open holographic systems with defects to certain semiclassical Virasoro blocks. The Heun-type correlators there look suspiciously similar to the normal modes of the non-BPS Janus \cite{gutperle2025januscorrelatorsheunsequation}, and coupling them together indicates similar expressions for the quasinormal modes. The WKB limit was also discussed extensively in these languages in \cite{jia2026exactholographicthermalspectral}. Following \cite{argüelles2026scaleinvariantopenquantumsystems}, we are also testing the universal scaling laws arise from thermal spectral functions in presence of a conformal bath in the double-trace coupling between AdS and holographic baths \cite{future}.

Last but not least, one may investigate whether the relation between BOPE data and ultraviolet dissipation extends beyond holography. In particular,  boundary and interface bootstrap methods provide a framework for constraining defect dimensions and BOPE coefficients through crossing symmetry \cite{Liendo_2013,Gliozzi_2015,MazacRastelliZhou:2019AnalyticBCFT}. The Tauberian theorem  already makes some progress but is not enough in determining the density of states and individual BOPE coefficients separately. It is conceivable that these dOPE data can be studied analytically in the large-$n$ fixed-$k$ limit considered here.

\acknowledgments
The author would like to thank Andreas Karch for some of the initial ideas of this work, and Jiaxin Qiao for helpful discussions. Certain calculations and writing of this manuscript were done with the help of OpenAI's GPT-5.6 Sol and GPT-6 Astra. This work was supported in part by the U.S. Department of Energy under Grant No. DE-SC0022021 and a grant from the Simons Foundation (Grant 651678, AK).

\appendix

\section{Pure-\texorpdfstring{$\AdS_3$}{AdS3} BOPE coefficients}
\label{app:pure-ads}

Here we check the normalizations in Section \ref{sec:closed-bope} for residues of normal modes and the BOPE coefficients in pure $\AdS_3$.  For
$A(x)=\cosh x$, the optical coordinate and warp factor are
$u=2\arctan e^x$ and $A(u)=\csc u$, so $L=\pi$ and $\rho=1$.
Writing $\nu=\Delta-1>0$, the Liouville potential is
\begin{equation}
  V_\Delta(u)=\left(\nu^2-\frac14\right)\csc^2u-\frac14.
\end{equation}
The eigenfunction with unit normalizable coefficient at the left endpoint
is
\begin{equation}
  \Psi_L(u;q_n)=
  (\sin u)^{\nu+\frac12}
  \frac{C_n^{(\nu+\frac12)}(\cos u)}
       {C_n^{(\nu+\frac12)}(1)}.
\end{equation}
Its eigenvalue is $q_n^2=(n+\Delta)(n+\Delta-1)$, hence
$d_n=n+\Delta$ and $\Omega_{n,k}=n+\Delta+k$ exactly.
Gegenbauer orthogonality gives the exact endpoint-normalized norm
\begin{equation}
  \mathcal I_n^{\AdS}
  =\frac{2^{2\nu+1}\Gamma(\nu+1)^2}{2d_n-1}
    \frac{\Gamma(n+1)}{\Gamma(n+1+2\nu)}.
  \label{eq:pure-ads-exact-norm}
\end{equation}
Equation~\eqref{eq:real-axis-norming-residue} then gives
\begin{equation}
  \mathcal R_n^{\AdS}
  =\frac{2}{4^\nu\Gamma(\nu)^2}
    \frac{\Gamma(n+1+2\nu)}{\Gamma(n+1)}.
  \label{eq:pure-ads-exact-residue}
\end{equation}
This weight is positive and independent of the descendant label $k$.

The projected response also checks the residue directly.  For
noninteger $\nu$, set $d=\omega-k$ as in
Section~\ref{sec:closed-response}.  A solution normalizable at $u=\pi$
is proportional to
\begin{equation*}
  (\sin u)^{\nu+\frac12}
  {}_2F_1\!\left(\nu+1-d,\nu+d;\nu+1;
                        \frac{1+\cos u}{2}\right).
\end{equation*}
Its connection coefficients at $u=0$ give
\begin{equation*}
  \frac{\beta_k^{\AdS}}{\alpha_k^{\AdS}}
  =\frac{\Gamma(-\nu)}{4^\nu\Gamma(\nu)}
    \frac{\Gamma(\nu+1-d)\Gamma(\nu+d)}
         {\Gamma(1-d)\Gamma(d)}.
\end{equation*}
Using $G_R^{\mathrm{proj}}=2\nu\beta/\alpha$ and the Gamma reflection
formula gives
\begin{align}
  G_{R,k}^{\AdS,\mathrm{proj},(\nu)}(\omega)
  ={}&-\frac{2\pi}{4^\nu\Gamma(\nu)^2}
       \frac{\Gamma(d+\nu)}{\Gamma(d-\nu)}
       \left[\cot\!\bigl(\pi(d-\Delta)\bigr)+\cot(\pi\nu)\right]
       \nonumber\\
  &+G_{\mathrm{loc}}^{\AdS}(\omega).
  \label{eq:pure-ads-exact-response}
\end{align}
Here $G_{\mathrm{loc}}^{\AdS}$ is the local contact term.  At
$d=d_n$, the negative residue agrees with
\eqref{eq:pure-ads-exact-residue}.  Also,
\begin{equation*}
  \frac{\Gamma(d+\nu)}{\Gamma(d-\nu)}
  =[d(d-1)]^\nu[1+O(d^{-2})],
  \qquad d\to+\infty,
\end{equation*}
which reproduces \eqref{eq:system-response-leading} and
\eqref{eq:massive-bope-WKB} at $\rho=1$.  At positive integer $\nu$,
the response requires subtraction of a divergent local term, but the
norms and residues above remain valid.

For the scalar action \eqref{eq:scalar-action}, the ambient two-point
normalization is
\begin{equation*}
  \mathcal A_{\mathbf1}
  =\frac{(2\Delta-2)\Gamma(\Delta)}{\pi\Gamma(\Delta-1)}
  =\frac{2(\Delta-1)^2}{\pi}.
\end{equation*}
In the reduced-correlator convention of
Section~\ref{sec:massive-tauberian}, the trivial interface has
$\mathcal F_{\AdS}(\xi)=\mathcal A_{\mathbf1}\xi^{-\Delta}$.
Its defect-block expansion has BOPE coefficients
\begin{equation}
  b_{\AdS,n}^2
  =\frac{2(\Delta-1)^2}{\pi}
    \frac{(\Delta)_n^2}{n!(2\Delta+n-1)_n},
  \qquad (a)_n=\frac{\Gamma(a+n)}{\Gamma(a)}.
  \label{eq:pure-ads-exact-bope}
\end{equation}
Direct substitution gives
$\mathcal K_\Delta(n+\Delta)b_{\AdS,n}^2=\mathcal R_n^{\AdS}$,
checking \eqref{eq:massive-bope-dictionary}.
The same coefficients also obey
\begin{equation*}
  \mathfrak w(n+\Delta)b_{\AdS,n}^2
  =\frac{4(2n+2\Delta-1)}{\pi\Gamma(\Delta-1)^2}
    \frac{\Gamma(n+2\Delta-1)}{\Gamma(n+1)}
  \sim\frac{8n^{2\Delta-1}}{\pi\Gamma(\Delta-1)^2}.
\end{equation*}
Summing up to $n+\Delta\leq\Lambda$ reproduces the cumulative
coefficient in \eqref{eq:massive-tauberian-cumulative}.

\section{1D conformal bootstrap from Tauberian theorem}
\label{app:tauberian}

Let us perform a consistency check this result using the $SL(2,\mathbb{R})$ bootstrap of 1D conformal blocks. We follow the arguments of \cite{QiaoRychkov2017} closely.  For Euclidean points
$\mathbf x_i=(\tau_i,x_{\perp i})$ on the same side, use
\begin{equation*}
  \xi=\frac{(\tau_1-\tau_2)^2+(x_{\perp1}-x_{\perp2})^2}
              {4x_{\perp1}x_{\perp2}},
  \qquad
  \langle\mathcal O_S(\mathbf x_1)\mathcal O_S(\mathbf x_2)\rangle_I
  =\frac{\mathcal F_I(\xi)}
  {(2x_{\perp1})^\Delta(2x_{\perp2})^\Delta}.
\end{equation*}
The defect expansion takes the form
\begin{equation}
  \mathcal F_I(\xi)=b_{I,\mathbf1}^2
      +\int_{(0,\infty)}\widehat g_d(\xi)\,\dd\mu_I(d),
  \qquad
  \widehat g_d(\xi)=\xi^{-d}
       {}_2F_1\!\left(d,d;2d;-\frac1\xi\right),
  \label{eq:defect-block}
\end{equation}
with the positive primary measure
\begin{equation}
  \dd\mu_I(d)=\sum_{a:\,d_{I,a}>0}
       b_{I,a}^2\delta(d-d_{I,a})\,\dd d.
  \label{eq:massive-tauberian-measure}
\end{equation}
Here $a$ includes degeneracies, and a possible defect identity has been
displayed separately.  Positivity holds for identical Hermitian operators
on the same side; convergence of this OPE is discussed in
\cite{LauriaMeineriTrevisani:2018RadialDefect}.

Set
\begin{equation}
  z=\frac1{1+\xi},\qquad \varepsilon=1-z=\frac{\xi}{1+\xi}.
  \label{eq:defect-block-variables}
\end{equation}
Pfaff's transformation identifies the defect block with the
one-dimensional global conformal block:
\begin{equation}
  \widehat g_d(\xi)=G_d(z),\qquad
  G_d(z)=z^d{}_2F_1(d,d;2d;z).
  \label{eq:defect-block-sl2}
\end{equation}
As $\xi\to0^+$, the  identity dominates the bulk OPE and we have
\begin{equation}
  \mathcal F_I(\xi)=\mathcal A_{\mathbf1}\xi^{-\Delta}[1+o(1)]
  \qquad(\xi\to0^+).
  \label{eq:massive-tauberian-identity}
\end{equation}
Since $\varepsilon/\xi\to1$, the positive-block Tauberian theorem of
\cite{QiaoRychkov2017} applies with external dimension $\Delta/2$ in
that theorem's notation.  Define
\begin{equation}
  \mathfrak w(d)=\frac{2\Gamma(2d)}{\Gamma(d)^2},\qquad
  Q_I(\Lambda)=\int_{(0,\Lambda]}\mathfrak w(d)\,\dd\mu_I(d).
  \label{eq:massive-tauberian-Q}
\end{equation}
The exact weight satisfies
$\mathfrak w(d)=4^d\sqrt{d/\pi}[1+O(d^{-1})]$, so positivity allows
it to replace the asymptotic weight in the theorem without changing
the leading coefficient.  The result of the integral is
\begin{equation}
  Q_I(\Lambda)=\frac{2\mathcal A_{\mathbf1}}
  {\Delta\Gamma(\Delta)^2}\Lambda^{2\Delta}[1+o(1)].
  \label{eq:massive-tauberian-general}
\end{equation}
The coefficient can be checked from the large-dimension block kernel
$\mathfrak w(d)K_0(2d\sqrt\varepsilon)$ and the integral
$\int_0^\infty t^{2\Delta-1}K_0(2t)\,\dd t
=\Gamma(\Delta)^2/4$.  For the canonical scalar action
\eqref{eq:scalar-action}, $\mathcal A_{\mathbf1}=2(\Delta-1)^2/\pi$,
as checked in appendix~\ref{app:pure-ads}.  Thus
\begin{equation}
  Q_I(\Lambda)=\frac{4}{\pi\Delta\Gamma(\Delta-1)^2}
  \Lambda^{2\Delta}[1+o(1)].
  \label{eq:massive-tauberian-cumulative}
\end{equation}

The connection to the projected residues follows from the exact identity
\begin{equation}
  \mathcal K_\Delta(d)
  =\frac{\pi2^{1-2\Delta}}{2d-1}\mathfrak w(d).
  \label{eq:massive-K-w}
\end{equation}
For a fixed cutoff $d_*>1/2$, Stieltjes integration by parts gives
\begin{equation*}
  \int_{(d_*,\Lambda]}\frac{\dd Q_I(d)}{2d-1}
  =\frac{Q_I(\Lambda)}{2\Lambda-1}
    +\int_{d_*}^{\Lambda}\frac{2Q_I(d)}{(2d-1)^2}\,\dd d+O(1).
\end{equation*}
Substituting \eqref{eq:massive-tauberian-cumulative} therefore yields
\begin{equation}
  \sum_{d_*<d_{I,a}\leq\Lambda}
    \mathcal K_\Delta(d_{I,a})b_{I,a}^2
  =\frac{2^{3-2\Delta}}
  {(2\Delta-1)\Gamma(\Delta-1)^2}
  \Lambda^{2\Delta-1}[1+o(1)].
  \label{eq:massive-tauberian-pole-sum}
\end{equation}
If the selected radial tower exhausts the scalar correlator at classical
probe order, every summand is $\mathcal R_n^S$.  Its spacing tends to
$1/\rho$, while \eqref{eq:massive-bope-WKB} gives
$\mathcal R_n^S\sim2^{3-2\Delta}
d_n^{2\Delta-2}/[\Gamma(\Delta-1)^2\rho]$.
Summing these weights reproduces \eqref{eq:massive-tauberian-pole-sum}:
the factor $\rho$ counting states cancels the inverse density in each
residue. Hence the above response residues and the constants $\mathcal{K}_\Delta$ is consistent with the Tauberian theorem for 1D conformal blocks at large defect dimension\footnote{We thank GPT-Astra for bringing up this consistency check and helped with its calculations.}.

\section{Proofs for the geometric bound}
\label{app:geometric-proof}

In this appendix we supplement the geometric proofs used in
Section~\ref{sec:new-bound}, including the upper bound of $I_2$ and the saturation condition.

\subsection{Upper bound on $I_2$}

Let $u_-$ and $u_+$ be the first and last points at which $B=M$.
Measure distance outward from each of these points and define
\begin{equation*}
  D_-(\zeta)=M-B(u_--\zeta),\quad 0\leq\zeta\leq u_-;
  \qquad
  D_+(\zeta)=M-B(u_++\zeta),\quad 0\leq\zeta\leq L-u_+.
\end{equation*}
For either cap $[0,\ell]$, we have
$D(0)=D'(0)=0$, $D(\ell)=M$, and $0\leq D\leq M$.
The middle interval adds a nonnegative deficit.  The NEC gives
\begin{equation}
  D''\leq M-D.
  \label{eq:bound-deficit-inequality}
\end{equation}

Consider any connected component of $\{D'>0\}$.  At its left endpoint
$D'=0$; denote the deficit there by $D_0\in[0,M]$.  Multiplying
\eqref{eq:bound-deficit-inequality} by $D'$ and integrating within
the component gives
\begin{equation*}
  \frac12D'^2
  \leq MD-\frac12D^2-\left(MD_0-\frac12D_0^2\right).
\end{equation*}
The subtracted term is nonnegative, hence
\begin{equation}
  D'^2\leq2MD-D^2
  \qquad\hbox{wherever }D'>0.
  \label{eq:bound-deficit-energy}
\end{equation}
Continuity of $D$ and $D'$ preserves this estimate across the joins.

To include decreasing and flat pieces, define
\begin{equation*}
  H(z)=\frac{z}{\sqrt{2Mz-z^2}},\quad 0<z\leq M,
  \qquad H(0)=0.
\end{equation*}
Equation~\eqref{eq:bound-deficit-energy} gives $H(D)D'\leq D$ when
$D'>0$, and the inequality also holds when $D'\leq0$.  Thus
\begin{equation}
  \int_0^\ell D(\zeta)\,\dd\zeta
  \geq\int_0^\ell H(D(\zeta))D'(\zeta)\,\dd\zeta
  =\int_0^M\frac{z\,\dd z}{\sqrt{2Mz-z^2}}
  =M\left(\frac\pi2-1\right).
  \label{eq:bound-single-cap}
\end{equation}
The last integral uses $z=M(1-\cos\theta)$.  Adding the two caps and
the nonnegative middle deficit proves \eqref{eq:bound-cap-deficit},
or equivalently \eqref{eq:bound-I2-sharp}.

\subsection{Equality conditions and supporting matter}

The saturation of \eqref{eq:bound-main} first requires zero middle deficit $D$, so $B=M$ throughout
$[u_-,u_+]$.  On either cap, the continuous nonnegative function
$D-H(D)D'$ must vanish.  By the definitions of $u_-$ and $u_+$,
$D(\zeta)>0$ for $0<\zeta\leq\ell$.  It follows that
$D'=\sqrt{2MD-D^2}$ there, excluding decreasing or flat pieces.
The solution is
\begin{equation}
  D(\zeta)=M(1-\cos\zeta),
  \qquad 0\leq\zeta\leq\frac\pi2.
  \label{eq:bound-equality-deficit}
\end{equation}
There is no initial interval with $D=0$ by the definitions of
$u_-$ and $u_+$.  The condition $D(\ell)=M$ fixes $\ell=\pi/2$,
and $D'(\ell)=1$ fixes $M=1$.
The complete equality family is therefore
\begin{equation}
  B_{\mathrm{sat}}(u)=
  \begin{cases}
    \sin u,
      &0\leq u\leq\dfrac\pi2,\\[4pt]
    1,
      &\dfrac\pi2\leq u\leq L-\dfrac\pi2,\\[4pt]
    \sin(L-u),
      &L-\dfrac\pi2\leq u\leq L.
  \end{cases}
  \label{eq:bound-saturating-profile}
\end{equation}
Its integral is $I_2=L-\pi+2$.  Transforming back to $x$ gives
\eqref{eq:bound-saturating-warp-factor}.

Let us also verify that it has positive local energy and does not behave like a thin brane.
The independent components of the matter stress tensor $8\pi G_N T_{\mu\nu}=G_{\mu\nu}-g_{\mu\nu}$ from \eqref{eq:closed-metric} are
\begin{equation}
  8\pi G_NT^{\mathrm{matt}}_{xx}
    =\frac{1+(A')^2}{A^2}-1,
  \qquad
  8\pi G_NT^{\mathrm{matt}}_{ij}
    =\left(\frac{A''}{A}-1\right)g_{ij},
  \label{eq:bound-saturating-stress}
\end{equation}
where $i,j$ are tangent to the $\AdS_2$ slices.  Substitution of
\eqref{eq:bound-saturating-warp-factor} gives zero stress on the caps.
In the throat it gives
$(T^{\mathrm{matt}})^x{}_x=0$ and
$(T^{\mathrm{matt}})^i{}_j=-\delta^i{}_j/(8\pi G_N)$.
The energy density is
$-(T^{\mathrm{supp}})^t{}_t=1/(8\pi G_N)>0$.
Since $A$ and $A'$ are continuous, the joins have no delta-function layer.

\section{Janus interfaces}
\label{app:janus-examples}

In this appendix we check our claims against commonly used examples like Janus \cite{Bak_2007} and BPS-Janus interfaces \cite{Chiodaroli_2010}. In particular we check the bound assumptions on the warpfactor for our bottom-up models  used in
Section~\ref{sec:closed-optical}, then evaluate $c_{UV}$ and check the bounds in Section~\ref{sec:new-bound}
for these examples.

\subsection{The mild bound on $A(x)$}
\label{app:endpoint-model-checks}

We now check \eqref{eq:assumption1} in the inward optical distance
$y$.  For Einstein--scalar Janus \cite{Bak_2007}, $B=1/A_j$ obeys
\begin{equation*}
  (B_y)^2=1-B^2+\frac{j^2}{4}B^4,
  \qquad B_{yy}=-B+\frac{j^2}{2}B^3,
  \qquad B(0)=0,\quad B_y(0)=1.
\end{equation*}
The analytic solution near $B=0$ gives
\begin{equation*}
  A_j(y)=\frac1y+\frac y6+\frac{7-9j^2}{360}y^3+O(y^5).
\end{equation*}
Consequently, for $\varepsilon_j=yA_j-1$,
\begin{equation*}
  \varepsilon_j=\frac{y^2}{6}+O(y^4),\qquad
  (\varepsilon_j)_y=\frac y3+O(y^3),\qquad
  (\varepsilon_j)_{yy}=\frac13+O(y^2).
\end{equation*}
These estimates hold at both ends for each fixed $j<1$.
As $j\to1$, the centered warp factor tends to a constant and the
two AdS ends recede to infinite optical distance.  The limiting
geometry is outside the finite-interval class.

For the massless internal singlet of D1/D5 BPS-Janus, the separated
equation of \cite{Chiodaroli:2016jod} reads
\begin{equation*}
  \frac{\dd}{\dd x}\left(\cosh^2x\,\frac{\dd f}{\dd x}\right)
       +\kappa^2q^2f=0.
\end{equation*}
It is the canonical massless radial equation for
$A_{\rm eff}(x)=\cosh x/\kappa$.  This statement concerns the selected
singlet sector; other internal modes can have position-dependent
effective masses.  Since the inward optical distance is
$y=2\kappa\arctan e^{-|x|}$,
\begin{equation*}
  A_{\rm eff}(y)=\frac1{\kappa\sin(y/\kappa)}
  =\frac1y+\frac{y}{6\kappa^2}
      +\frac{7y^3}{360\kappa^4}+O(y^5).
\end{equation*}
This analytic expansion satisfies \eqref{eq:assumption1}.  In fact,
\begin{equation*}
  V_{\Delta=2}(u)
  =\frac{3}{4\kappa^2}\csc^2(u/\kappa)-\frac1{4\kappa^2},
\end{equation*}
 so the remainder relative to \eqref{eq:bounded-endpoint-remainder}
is exactly $-1/(4\kappa^2)$.  The three-dimensional
$\mathcal N=8$ Janus metric of \cite{Chen_2021} has the
same scaled-cosh warp factor, with $\kappa=\cosh Q$ in terms of its
deformation parameter $Q$, and satisfies the same endpoint estimates.

However, there are examples where these assumptions fail. For example,
truncation~1 of \cite{Chen_2022}, in units
with asymptotic AdS radius one, has
\begin{equation*}
  A(x)=\cosh x-b,\qquad b=\tanh^2(Q_0/2),\qquad 0<b<1,
\end{equation*}
where $Q_0$ is the scalar value at the turning point.  Its inverse
warp factor obeys $B_{yy}=b+(b^2-1)B$, with $B(0)=0$ and $B_y(0)=1$.
Thus
\begin{equation*}
  A(y)=\frac1y-\frac b2+\frac{b^2+2}{12}y+O(y^2),
  \qquad \varepsilon(y)=-\frac b2y+O(y^2).
\end{equation*}
Although the metric is smooth and asymptotically AdS at both ends, the assumptions
$\varepsilon=O(y^2)$ and $\varepsilon_y=O(y)$ fail for this case.  Substitution
into the Liouville potential gives
\begin{equation*}
  V_\Delta(y)=\frac{\nu^2-\frac14}{y^2}
          +\frac{b(\frac54-\nu^2)}{y}+O(1).
\end{equation*}
The remainder after subtracting the inverse-square term is unbounded unless
$\nu^2=5/4$. We leave the estimate of defect spectra and normal modes for this example to future work.

If we furthermore allow non-smooth solutions, more subtleties arise. For example, thin-brane models have pure-AdS caps
but distributional terms in the interior potential
\cite{BachasEtAl:2023ThickBranes}.

\subsection{Non-BPS Janus}
\label{sec:janus-optical}

Let us review the normal modes of non-BPS Janus following \cite{gutperle2025januscorrelatorsheunsequation}. After translating the transverse coordinate, the
warpfactor is
\begin{equation}
  A_j(x)^2=\frac{1+a\cosh 2x}{2},
  \qquad a=\sqrt{1-j^2},\qquad 0\leq j<1.
  \label{eq:janus-warp}
\end{equation}
Here $j=|c_J|$ in the conventions of
\cite{gutperle2025januscorrelatorsheunsequation}, and the magnitude of the
scalar jump is $\log[(1+j)/(1-j)]$.  The radial equation is of Heun
type, but the optical length follows directly by setting $t=\tanh x$:
\begin{align}
  L_j
  &=\sqrt{\frac{2}{1+a}}
    \int_{-1}^{1}\frac{\dd t}
    {\sqrt{(1-t^2)(1-\eta t^2)}}
    =2\sqrt{\frac{2}{1+a}}\,K(\eta),
    \qquad \eta=\frac{1-a}{1+a},
  \label{eq:janus-optical-length}
\end{align}
where the complete elliptic integral uses the parameter convention
\begin{equation*}
  K(\eta)=\int_0^{\pi/2}
    (1-\eta\sin^2\vartheta)^{-1/2}\dd\vartheta.
\end{equation*}
The endpoint conditions \eqref{eq:assumption1} hold for every fixed
$j<1$, as verified below.  Thus
\begin{equation}
  \rho_j=\frac{2}{\pi}\sqrt{\frac{2}{1+a}}K(\eta),
  \qquad
  d_n=\frac12+\frac{n+\Delta-\frac12}{\rho_j}+O(n^{-1}).
  \label{eq:janus-optical-spectrum}
\end{equation}
The stable standard-$\AdS_2$ branch also requires $q_0^2>-1/4$;
this low-level condition does not follow from the asymptotic density.

Expanding the elliptic integral at small deformation gives
\begin{equation}
  \rho_j=1+\frac{3j^2}{16}+O(j^4).
  \label{eq:janus-density-small-jump}
\end{equation}
The coefficient of $j^2$ in $d_n$ at large $n$ is therefore
$-\frac3{16}(n+\Delta-\frac12)+O(n^{-1})$.  This agrees with the
large-level expansion of the perturbative Heun spectrum in
\cite{gutperle2025januscorrelatorsheunsequation}, including its constant
term. 

\subsection{$c_{UV}$ and the bound}
\label{sec:janus-charge-checks}

We can write down the three central-charges like quantities for non-BPS Janus from \eqref{eq:janus-warp}. The minimum warpfactor and the second
geometric integral are
\begin{equation}
  \Amin=\sqrt{\frac{1+a}{2}},\qquad
  I_2=2\int_{-1}^{1}\frac{\dd t}{(1+a)-(1-a)t^2}
      =\frac{2\operatorname{arctanh}j}{j},
  \label{eq:janus-geometric-data}
\end{equation}
with the $j=0$ value defined by continuity.  Thus 
\begin{equation}
  \frac{\cUV}{c}=\frac{\pi}{2K(\eta)}\sqrt{\frac{1+a}{2}},
  \qquad
  \frac{\ceff}{c}=\sqrt{\frac{1+a}{2}},
  \qquad
  \frac{\cLR}{c}=\frac{j}{\operatorname{arctanh}j}.
  \label{eq:janus-three-charges}
\end{equation}
One can check that \eqref{eq:final-bound} holds for every $0\le j\le 1$.

For comparison, let us recall BPS-Janus with $\kappa=\cosh\psi\cosh\theta\geq1$ being its deformation parameter.  The massless internal singlet has
the exact defect-primary spectrum \cite{Chiodaroli:2016jod}
\begin{equation}
  d_n^{(\kappa)}
  =\frac12+\sqrt{\frac14+\frac{(n+1)(n+2)}{\kappa^2}},
  \qquad n=0,1,\ldots.
  \label{eq:closed-bps-janus-spectrum}
\end{equation}
Therefore
\begin{equation}
  \rho_{\rm BPS}=\kappa.
  \label{eq:closed-bps-janus-density}
\end{equation}
The separated massless equation has effective warp factor
$A_{\rm eff}(x)=\cosh x/\kappa$, as checked below, and hence optical
length $L=\pi\kappa$.  The independently calculated super-Janus charges
are \cite{Baig:2024hfc,karch2026universalitydissipationholographicinterfaces}
\begin{equation*}
  \frac{\cUV}{c}=\frac{\ceff}{c}=\frac1\kappa,
  \qquad \frac{\cLR}{c}=\frac1{\kappa^2}.
\end{equation*}
  Thus it also satisfies the bound. In addition, $\cUV=\ceff$ is special to this case.

\bibliographystyle{JHEP}
\bibliography{references_new.bib,tauberian_refs.bib}

\end{document}